\documentclass[english,11pt]{article}
\usepackage{geometry}                
\usepackage{graphicx}
\usepackage{amssymb}
\usepackage{amsmath,amssymb}
\usepackage{color}
\usepackage{float}

\definecolor{darkblue}{rgb}{0,0,0.5}
\usepackage[colorlinks,linkcolor=black,citecolor=darkblue,urlcolor=darkblue]{hyperref}

\newcommand{\be}{\begin{equation}}
\newcommand{\ee}{\end{equation}}
\newcommand{\bea}{\begin{eqnarray}}
\newcommand{\eea}{\end{eqnarray}}

\allowdisplaybreaks

\begin{document}  \thispagestyle{empty}

\vskip-0.5cm
\begin{flushright}
FERMILAB-PUB-11-330-T \\ EFI-11-21 
\end{flushright}
\vskip0.5cm

\begin{center}
{\Large \bf \boldmath 
Signals of CP Violation Beyond the MSSM \\ in Higgs and Flavor Physics
}

\vskip1.5cm
{\bf Wolfgang~Altmannshofer$^{a}$, Marcela Carena$^{a,b}$, Stefania~Gori$^{b}$ \\ and Alejandro~de~la~Puente$^{a,c}$}
%
%
\vskip0.8cm
$^a$ {\em Fermi National Accelerator Laboratory, P.O. Box 500, Batavia, IL 60510, USA}\\[6pt]
$^b$ {\em Enrico Fermi Institute, University of Chicago, Chicago, IL 60637, USA}\\[6pt]
$^c$ {\em Department of Physics, University of Notre Dame, Notre Dame, IN 46556, USA}

\vskip1.5cm

\end{center}
\begin{abstract}
We study an extension of the Higgs sector of the Minimal Supersymmetric Standard Model (MSSM), considering the effects of new degrees of freedom at the TeV scale, and allowing for sources of CP violation beyond the MSSM (BMSSM). We analyze the impact of the BMSSM sources of CP violation on the Higgs collider phenomenology and on low energy flavor and CP violating observables.
We identify distinct Higgs collider signatures that cannot be realized, either in the case without CP violating phases or in the CP violating MSSM, and investigate the prospects to probe them at the Tevatron and the LHC. The most striking benchmark scenario has three neutral Higgs bosons that all decay dominantly into $W$ boson pairs and that are well within the reach of the 7~TeV LHC run. On the other hand, we also present scenarios with three Higgs bosons that have masses $M_{H_i} \gtrsim 150$~GeV and decay dominantly into $b \bar b$. Such scenarios are much more challenging to probe and can even lie completely outside the reach of the 7~TeV LHC run.
We explore complementary scenarios with standard MSSM Higgs signals that allow to accommodate a non-standard $B_s$ mixing phase as indicated by D0, as well as the excess in $B_s \to \mu^+ \mu^-$ candidates recently reported by CDF. We find that, in contrast to the MSSM, a minimal flavor violating soft sector is sufficient to generate significant corrections to CP violating observables in meson mixing, compatible with EDM constraints. 
In particular, a $B_s$ mixing phase $S_{\psi\phi} \lesssim 0.15$, can be achieved for specific regions of parameter space, compatible with all the presently available experimental constraints on flavor observables. Such a non-standard $B_s$ mixing phase would unambiguously imply a sizable suppression of $S_{\psi K_S}$ with respect to the SM prediction and a BR$(B_s \to \mu^+ \mu^-)$ close to its 95\%~C.L. upper bound of $1.1\times 10^{-8}$.
\end{abstract}

\newpage
\tableofcontents
\section{Introduction}

Extensions of the Standard Model (SM) that are based on Supersymmetry (SUSY) are arguably the most popular models of New Physics (NP) at the TeV scale. In particular the Minimal Supersymmetric Standard Model (MSSM) is one of the most studied NP models. On the other hand, arguments based on the naturalness criterion suggest that the minimal Higgs sector of the MSSM should be extended in order to raise the tree level mass of the lightest Higgs boson above its MSSM limit of $M_Z \cos2\beta$~\cite{Casas:2003jx,Barbieri:2006dq,Cassel:2009ps}. Many such Supersymmetric models that contain physics Beyond the MSSM (e.g. an enlarged gauge sector, additional singlets) have been constructed~(see e.g. \cite{Batra:2003nj,Harnik:2003rs,Birkedal:2004zx,Maloney:2004rc,Barbieri:2006bg,Martin:2009bg,Lodone:2010kt,Franceschini:2010qz,Delgado:2010uj,Delgado:2010cw,Bertuzzo:2011ij}). As long as the scale of the beyond MSSM physics $M$ is sufficiently larger than the mass scales of the MSSM Higgs sector, it is possible to describe the effects of the new degrees of freedom in an effective theory approach, where higher dimensional operators are added to the MSSM~\cite{Brignole:2003cm,Dine:2007xi,Antoniadis:2007xc,Antoniadis:2008es,Carena:2009gx,Antoniadis:2009rn,Antoniadis:2010nb}. It was shown that in such an effective field theory approach (the so called BMSSM), the lightest Higgs boson can easily be enhanced at the tree level and reach a mass of around 200~GeV, as long as $M$ is not larger than a few TeV.

In this work we consider the BMSSM framework where the MSSM Higgs sector is extended by the leading SUSY preserving and SUSY breaking dimension 5 operators~\cite{Dine:2007xi}. 
Extensions of the MSSM including also dimension 6 operators have been studied in \cite{Carena:2009gx,Antoniadis:2009rn,Antoniadis:2010nb} and examples of possible UV completions have been presented in~\cite{Carena:2009gx}. One interesting feature of the model in~\cite{Dine:2007xi} is, that in contrast to the MSSM it allows for CP violation in the Higgs sector already at the tree level.
Most studies of this framework in the literature assume the absence of new sources of CP violation and consist of analyses of the vacuum structure of the model~\cite{Batra:2008rc,Blum:2009na}, dark matter~\cite{Bernal:2009hd,Bernal:2009jc,Berg:2009mq} and the Higgs collider phenomenology~\cite{Carena:2010cs}.  
The possible effects of CP violation induced by the higher dimensional operators have been mainly studied in the context of electro-weak baryogenesis~\cite{Blum:2008ym,Bernal:2009hd,Blum:2010by}.

Differently, in the first part of this work we study the impact of the BMSSM sources of CP violation on the Higgs collider phenomenology, extending the analysis of the CP conserving case of~\cite{Carena:2010cs}. In~\cite{Carena:2010cs} the expected signals of the CP conserving BMSSM at the Tevatron and at the LHC have been worked out in detail. Higgs production and decay patterns that are markedly different from the MSSM have been identified. One example are scenarios where both CP-even scalar Higgs bosons decay dominantly into a pair of gauge bosons.
In this work we identify characteristic collider signatures of the BMSSM with CP violation that cannot be realized, either in the case without CP violating phases or in the CP violating MSSM and investigate the prospects to probe the model at the Tevatron and the LHC.

In the second part of this work we analyze possible characteristic signals of the higher dimensional operators in flavor physics. As the higher dimensional operators mainly modify the spectrum and couplings of the neutral Higgs bosons of the MSSM, significant deviations from the MSSM predictions are expected in those flavor observables that are highly sensitive to the exchange of neutral Higgs bosons. Consequently, we analyze the rare $B_{d,s} \to \mu^+\mu^-$ decays as well as $B_{d,s} - \bar B_{d,s}$ mixing, that can receive sizable NP contributions in the large $\tan\beta$ regime from Higgs and double Higgs penguins, respectively. In particular, we study if the BMSSM with Minimal Flavor Violation (MFV), i.e. the BMSSM with no additional sources of flavor violation apart from the CKM matrix~\cite{Chivukula:1987py,Buras:2000dm,D'Ambrosio:2002ex} but new sources of CP violation from the dimension 5 operators, can accommodate a large phase in $B_s$ mixing as indicated by recent experimental results from Tevatron, especially the very recent D0 result on the like-sign dimuon charge asymmetry~\cite{Abazov:2011yk}.
In view of the excess of $B_s \to \mu^+ \mu^-$ candidates recently reported by CDF~\cite{Collaboration:2011fi}, we in particular also analyze the correlation between the $B_s$ mixing phase and the  BR($B_s\to\mu^+\mu^-$) in the BMSSM with MFV.
A complementary discussion of other flavor observables in the BMSSM appeared recently in~\cite{Bernal:2011pj}.

\bigskip

The paper is organized as follows. In Sec.~\ref{sec:BMSSM} we review the extended Higgs sector of the BMSSM with sources of CP violation at tree level. We discuss the Higgs potential in presence of the higher dimensional operators and the vacuum structure of the model. We analyze the Higgs spectrum emphasizing the possible role of the new CP violating phases.  
In Sec.~\ref{sec:EDMs} we discuss constraints coming from Electric Dipole Moments (EDMs) that are induced by the new phases appearing in the Higgs sector. The Higgs collider phenomenology of the model is discussed in section~\ref{sec:collider_pheno}. We address constraints coming from direct Higgs searches at LEP and Tevatron and present benchmark scenarios showing Higgs spectra and couplings that are specific to the BMSSM with CP violation. We outline the most promising Higgs search strategies in these scenarios. In Sec.~\ref{sec:Bsmixing} we discuss distinct BMSSM signals in flavor physics, concentrating on the phase of $B_s$ mixing and its correlation with the rare $B_s \to \mu^+\mu^-$ decay. We conclude in Sec.~\ref{sec:conclusions}.
The appendices contain some details on the chargino, neutralino and squark masses in the BMSSM, a short discussion about electroweak precision tests, as well as a compendium of loop functions.

\section{The Extended Higgs Sector of the Model}\label{sec:BMSSM}

We study the framework first presented in~\cite{Dine:2007xi}, where the leading higher dimensional operators are added to the MSSM Higgs sector. The scale $M$ at which these operators arise is assumed to be not far above the TeV scale.

\subsection{The Higgs Potential}

At the $1/M$ order, the most general Higgs superpotential reads~\cite{Dine:2007xi}

\begin{equation}\label{eq:operator1}
W=\mu \hat H_u \hat H_d+\frac{\omega}{2M}\left(\hat H_u \hat H_d\right)^2~,
\end{equation}
where $\hat H_u$ and $\hat H_d$ are the Higgs superfields with hypercharge $+1/2$ and $-1/2$, respectively and we denote $\hat H_u \hat H_d\equiv \hat H_u^+\hat H_d^--\hat H_u^0\hat H_d^0$. The dimensionless parameter $\omega$ is taken to be of order 1 and complex.

In addition to the $1/M$ suppressed term in the superpotential, a corresponding SUSY breaking term is added to the Lagrangian~\cite{Dine:2007xi}, 
\begin{equation}\label{eq:operator2}
\mathcal{L} \supset \alpha \frac{\omega m_S}{2 M} (H_u H_d)^2~,
\end{equation}
where $\alpha$ is another free parameter of order 1 and complex.
The scale $m_S$ is the scale of the SUSY breaking terms of the physics beyond the MSSM and therefore $m_S \ll M$ is necessary to integrate out the complete SUSY multiplets of the new degrees of freedom at a common scale M. The scale $m_S$ can be connected to the scale of the soft SUSY breaking terms of the MSSM (i.e. the sfermion and gaugino masses), but it is possible to allow for sizable differences between these scales. To ensure validity of the effective description of the BMSSM physics in terms of higher dimensional operators, $m_S$ (as well as $\mu$) has to be sufficiently small compared to $M$, for $\alpha$ and $\omega$ of O(1).
 
At the renormalizable level, the resulting tree level scalar potential then reads 
\begin{eqnarray} \label{eq:Higgs_potentialren} \nonumber
V_{\rm{ren}} &=& V_{\rm{MSSM}}+ \left( \alpha \frac{\omega m_S}{2 M} (H_u H_d)^2 - \frac{\omega \mu^*}{M} (H_u H_d)(H_u^\dagger H_u + H_d^\dagger H_d) ~+~ h.c.\right) \\
&=&(m_{H_u}^2 + |\mu|^2) H_u^\dagger H_u + (m_{H_d}^2 + |\mu|^2) H_d^\dagger H_d +\left( B\mu (H_u H_d) + h.c.\right) \nonumber \\
&& + \frac{g_2^2}{8 c_{W}} (H_d^\dagger H_d)^2 + \frac{g_2^2}{8 c_{W}}  (H_u^\dagger H_u)^2 - \frac{g_2^2}{4 c_{W}}  (H_d^\dagger H_d)(H_u^\dagger H_u) + \frac{g_2^2}{2}  (H_u^\dagger H_d)(H_d^\dagger H_u) \nonumber \\
&& + \left( \alpha \frac{\omega m_S}{2 M} (H_u H_d)^2 - \frac{\omega \mu^*}{M} (H_u H_d)(H_u^\dagger H_u + H_d^\dagger H_d) ~+~ h.c. \right)~,
\end{eqnarray}
and for later convenience we define
\begin{equation}\label{eq:lambda_treeBMSSM}
\lambda_5 = |\lambda_5| e^{i \phi_5} \equiv \frac{\alpha \omega m_S}{M} ~,~~~ \lambda_6 = |\lambda_6| e^{i \phi_6} \equiv \frac{\omega \mu^*}{M}~.
\end{equation}
The $1/M$ operator in the superpotential leads to two additional non-renormalizable dimension six terms
\begin{equation}\label{eq:Higgs_potentialnonren}
V_6 = \frac{\lambda_8}{M^2} (H_u H_d)(H_u^\dagger H_d^\dagger)(H_u^\dagger H_u) + \frac{\lambda_8^\prime}{M^2} (H_u H_d)(H_u^\dagger H_d^\dagger)(H_d^\dagger H_d)~,
\end{equation} 
with $\lambda_8 = \lambda_8^\prime = |\omega|^2$.
These terms are essential to stabilize the Supersymmetric electro-weak Symmetry Breaking (sEWSB) vacuae analyzed in~\cite{Batra:2008rc}. Thanks to these non-renor-malizable terms, the potential is automatically bounded from below.

At the $1/M^2$ order, there can be additional operators in the K\"ahler potential that modify the quartic couplings of the Higgs potential. Their possible impact has been analyzed in~\cite{Carena:2009gx,Antoniadis:2009rn}. In this work instead, we focus on the leading effects generated by the $1/M$ operators in~(\ref{eq:operator1}) and~(\ref{eq:operator2}). While additional $1/M^2$ operators can lead to an additional increase of the lightest Higgs mass, we do not expect them to change our main conclusions.

The three parameters $B\mu$, $\alpha$ and $\omega$ can in general be complex. 
We follow the usual convention adopted in studies of the MSSM and absorb the phase of $B\mu$ by a rephasing of the two Higgs doublets. In addition, we will assume all possible complex parameters of the MSSM (e.g. gaugino masses, $\mu$ parameter etc.) to be real. If these parameters were complex, CP violating effects in the Higgs sector would be possible at the 1-loop level and would add to the tree level effects. In this work we consider the phases of $\alpha$ and $\omega$ as the only Beyond the Standard Model sources of CP violation.

\subsection{The Minimum of the Potential}\label{sec:vacuum_stability}

We parametrize the Higgs fields as\footnote{Here we neglect the possibility of charge breaking vevs. As shown in~\cite{Blum:2009na} this is a good approximation in regions of parameter space that lead to a stable vacuum.}
\begin{equation}\label{eq:Higgs_fields}
H_u = e^{i\theta_u} \begin{pmatrix} H_u^+ \\ \frac{1}{\sqrt{2}} (v_u + h_u + i a_u) \end{pmatrix} ~, ~~~ H_d = e^{i\theta_d} \begin{pmatrix} \frac{1}{\sqrt{2}} (v_d + h_d + i a_d) \\ H_d^- \end{pmatrix}~,
\end{equation}
where $v_u = v \sin\beta = v s_\beta$ and $v_d = v \cos\beta = v c_\beta$ with $v = 246$~GeV are the two vacuum expectation values (VEVs) and $\tan\beta = t_\beta = v_u/v_d$ is their ratio.

While the relative phase of the Higgs fields $\theta_u - \theta_d$ can be rotated away by a $U(1)_Y$ transformation, $\theta\equiv\theta_u+\theta_d$ is a physical phase. Therefore, the following three extremal point conditions have to be satisfied at the minimum of the potential
\begin{equation}\label{eq:min}
\frac{\partial V}{\partial {\rm Re} H_u}=\frac{\partial V}{\partial {\rm Re} H_d}=\frac{\partial V}{\partial \theta}=0 ~.
\end{equation}
Using the first two conditions in Eq.~(\ref{eq:min}), the two soft masses $m_{H_u}$ and $m_{H_d}$ can be traded for $v$ and $\tan\beta$. 
The third condition determines the phase of the Higgs VEV, $\theta$, as a function of the phases of $\alpha$ and $\omega$ and reads
\begin{equation}\label{eq:mintheta}
v^2 c_\beta s_\beta |\lambda_5| \sin(\phi_5 + 2\theta) + v^2 |\lambda_6| \sin(\phi_6 + \theta) - 2 B\mu \sin \theta = 0 ~.
\end{equation}
Contrary to the MSSM, the BMSSM predicts in general already at tree level a non-zero phase of the Higgs fields at the minimum. In Sec.~\ref{sec:EDMs}, we will show that constraints coming from Electric Dipole Moments (EDMs) imply that the phase of the VEV is typically rather small $\theta \lesssim O(0.1)$.

\begin{figure}[t] \centering
\includegraphics[width=0.45\textwidth]{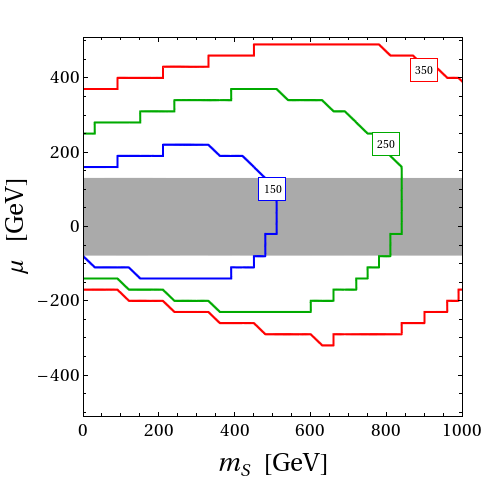} ~~~
\includegraphics[width=0.45\textwidth]{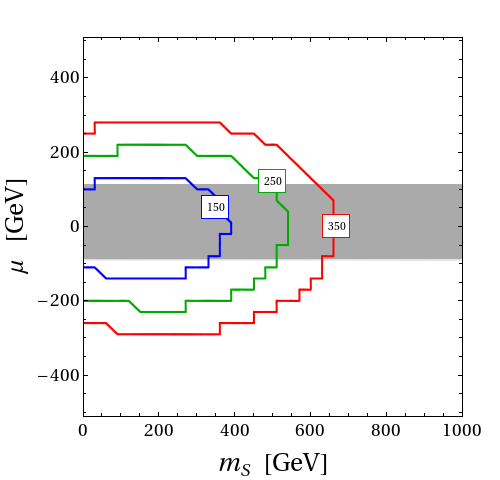}
\caption{\small 
Constraints in the $\mu-m_S$ plane from vacuum stability. The region within the blue, (green, red) contour leads to an absolute stable electroweak vacuum for a charged Higgs mass of 150~GeV (250~GeV, 350~GeV). The NP scale $M$ is fixed to 2~TeV, $\tan\beta=2$, $|\omega|=|\alpha|=1$, $m_{\tilde t}=800$~GeV and $A_t= 2 m_{\tilde t}$. In the left plot both $\alpha$ and $\omega$ are real and positive, while the right plot shows the situation with maximal phases Arg$(\alpha) = $ Arg$(\omega) = \pi/2$. 
The gray band is excluded by direct bounds on the chargino mass.
}
\label{fig:vacuum}
\end{figure}

To ensure that the stationary point of the potential defined by~(\ref{eq:min}) is a minimum, the corresponding Hessian has to be positive definite. However, the conditions to have a stationary point and the requirement on the Hessian do not necessarily lead to a unique solution. 
If the quartic couplings along the D-flat direction are negative, then a second minimum in the $v_u-v_d$ plane may arise for large field values that is stabilized by the $1/M^2$ suppressed terms in~(\ref{eq:Higgs_potentialnonren}). If this second minimum is deeper, the physical minimum at $v=246$~GeV is unstable and will decay into the second deeper minimum. In Fig.~\ref{fig:vacuum} we show in the $\mu-m_S$ plane the region that is allowed by the requirement of absolute vacuum stability. The criterion of absolute stability particularly leads to stringent {\it upper} bounds on the value of the $\mu$ parameter.
In accordance with the findings of~\cite{Blum:2009na}, we observe that the portion of allowed parameter space shrinks for smaller values of the charged Higgs mass. This behavior with $M_{H^\pm}$ holds, varying the value of the NP scale $M$ or the two phases of $\omega$ and $\alpha$, even if their value can have a rather sizable impact on the allowed values for $\mu$.
Combining the requirement of vacuum stability with the bound on the mass of the chargino, that leads to a {\it lower} bound on the absolute value of $\mu$ of $\sim 100$~GeV, the allowed regions for $\mu$ are strongly constrained, especially for low values of the charged Higgs mass (see the blue contours in Fig.~\ref{fig:vacuum} corresponding to $M_{H^\pm}=150$~GeV).

However, one should keep in mind that the requirement of absolute vacuum stability is rather conservative: it would be sufficient to impose that the EW vacuum is metastable, provided that its life time is longer than the age of the universe. This possibility has been discussed in~\cite{Blum:2009na} in the context of the BMSSM without CP violation.
Using simple analytic approximations for the bounce action~\cite{Duncan:1992ai}, we checked that the viable parameter space indeed opens up to some extent if we allow for a metastable vacuum. However, to be conservative, we require always absolute stability of the vacuum in the discussion of the Higgs phenomenology.

\subsection{The Higgs Spectrum}

We now briefly review the effects of the higher dimensional operators introduced in the previous section and in particular of the new physics phases on the Higgs spectrum. 
In order to keep a clear comparison to the case without CP violation, we write the mass matrices of the neutral Higgs bosons in the basis that would diagonalize it in absence of new sources of CP violation
\begin{equation}
\begin{pmatrix} h \\ H \end{pmatrix} = \begin{pmatrix} c_\alpha & -s_\alpha \\ s_\alpha & c_\alpha \end{pmatrix} \begin{pmatrix} h_u \\ h_d \end{pmatrix} ~, ~~~ \begin{pmatrix} G \\ A \end{pmatrix} = \begin{pmatrix} s_\beta & -c_\beta \\ c_\beta & s_\beta \end{pmatrix} \begin{pmatrix} a_u \\ a_d \end{pmatrix}~.
\end{equation}
The angle $\alpha$ is given by
\begin{eqnarray}\label{eq:alpha}
\sin2\alpha &=& - \frac{M_A^2 + M_Z^2}{M_H^2 - M_h^2} \sin2\beta + \frac{2 v^2 |\lambda_6| \cos(\phi_6 + 2\theta)}{M_H^2 - M_h^2} ~, \\ \label{eq:calpha}
\cos2\alpha &=& - \frac{M_A^2 - M_Z^2}{M_H^2 - M_h^2} \cos2\beta - \frac{v^2 |\lambda_5| \cos(\phi_5 + 2\theta)}{M_H^2 - M_h^2} \cos2\beta~.
\end{eqnarray}
In absence of CP violation, $M_A$ is the mass of the pseudoscalar Higgs\footnote{
In the CP violating scenario we are studying, $M_A$ as well as $M_h$ and $M_H$ are only auxiliary parameters and not physical masses.}
\begin{equation}
M_A^2 s_\beta c_\beta = B\mu \cos\theta - \frac{v^2}{2} |\lambda_6| \cos(\phi_6 + \theta) - v^2 |\lambda_5| s_\beta c_\beta \cos(\phi_5 + 2\theta)~,
\end{equation}
and $M_h^2$ and $M_H^2$ are the masses of the two scalars, namely the eigenvalues of the mass matrix
\begin{eqnarray}\label{eq:scalar_mass_matrix}
M_S^2 &=& M_A^2 \begin{pmatrix} c_\beta^2 & -c_\beta s_\beta \\ -c_\beta s_\beta & s_\beta^2 \end{pmatrix} + M_Z^2 \begin{pmatrix} s_\beta^2 & -c_\beta s_\beta \\ -c_\beta s_\beta & c_\beta^2 \end{pmatrix} \nonumber \\
&& + v^2 |\lambda_6| \cos(\phi_6 + \theta) \begin{pmatrix} 2 c_\beta s_\beta & 1 \\ 1 & 2 c_\beta s_\beta \end{pmatrix} + v^2 |\lambda_5| \cos(\phi_5 + 2\theta) \begin{pmatrix} c_\beta^2 & 0 \\ 0 & s_\beta^2 \end{pmatrix}~.
\end{eqnarray}
While the Goldstone boson $G$ is not affected by the presence of CP violation, the 3 physical Higgs bosons mix in presence of CP violating phases.
In the basis $(h,H,A)$, their mass matrix can be written as
\begin{equation}\label{eq:Higgs_mass_matrix}
\mathcal{M}_H^2 = \begin{pmatrix} 
M_h^2 & 0 & M_{hA}^2 \\ 
0 & M_H^2 & M_{HA}^2 \\ 
M_{hA}^2 & M_{HA}^2 & M_A^2
\end{pmatrix} ~,
\end{equation}
where the mixing terms are given by 
\begin{eqnarray}
M_{hA}^2 &=& - \frac{v^2}{2} \left( c_{\beta+\alpha} |\lambda_5| \sin(\phi_5 + 2\theta) - 2 s_{\beta-\alpha} |\lambda_6| \sin(\phi_6 + \theta) \right)~, \\
M_{HA}^2 &=& - \frac{v^2}{2} \left( s_{\beta+\alpha} |\lambda_5| \sin(\phi_5 + 2\theta) - 2 c_{\beta-\alpha} |\lambda_6| \sin(\phi_6 + \theta) \right)~.
\end{eqnarray}
The Higgs mass matrix (\ref{eq:Higgs_mass_matrix}) can be diagonalized by an orthogonal matrix $O$
\begin{equation}\label{eq:O}
O^T \mathcal{M}_H^2 O={\rm diag}(M_{H_1}^2,M_{H_2}^2,M_{H_3}^2) ~,
\end{equation}
where $M_{H_i}^2$ are the three eigenvalues.

\begin{figure}[t] \centering
\includegraphics[width=0.45\textwidth]{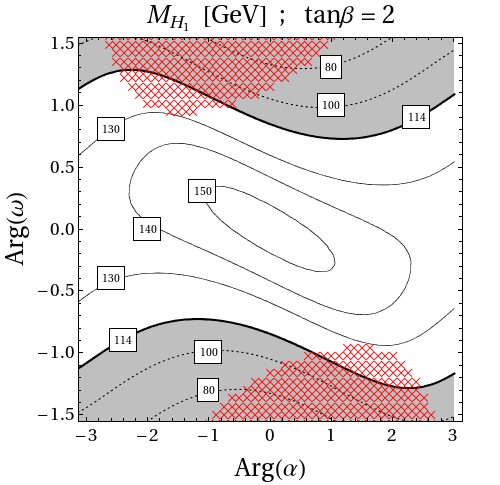} ~~~
\includegraphics[width=0.45\textwidth]{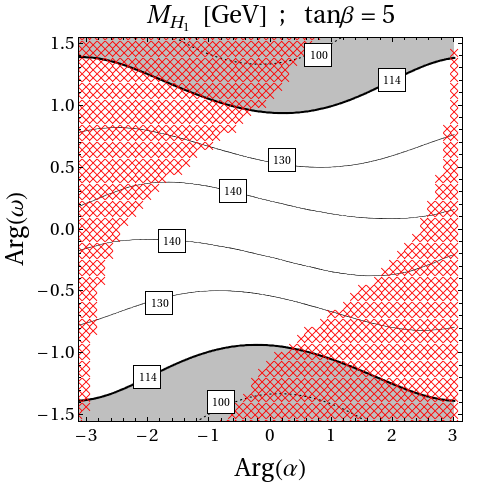}
\caption{\small 
The lightest neutral Higgs mass in the Arg$(\alpha)$ - Arg$(\omega)$ plane for two different values of $\tan\beta~=2,~5$. The remaining parameters are fixed to: $|\alpha| = |\omega| = 1$, $\mu = m_S = 150$~GeV, $M = 1.5$~TeV, $M_{H^\pm} = 200$~GeV, $m_{\tilde t} = 800$~GeV, $A_t = 2 m_{\tilde t}$. In the red hatched region the electro-weak symmetry breaking vacuum is only a local minimum of the potential.}
\label{fig:massh}
\end{figure}

In order to get an analytical understanding of the dependence of the Higgs spectrum of the model on the several phases, we give approximate expressions for the masses in the decoupling limit $M_A\gg M_Z$, performing a simultaneous expansion in $1/\tan\beta$ and $1/M$
\begin{eqnarray} \label{eq:Higgs_spectrum}
M_{H_1}^2 &\simeq& M_Z^2 + \frac{4v^2}{\tan\beta} |\lambda_6| \cos(\phi_6 + \theta) +\frac{v^4}{M_A^2} |\lambda_6|^2 \cos^2(\phi_6 + \theta) \nonumber \\
&& + \frac{3}{2\pi^2} \frac{m_t^4}{v^2} \left[ \log\left(\frac{\tilde m_t^2}{m_t^2}\right) + \frac{|A_t|^2}{\tilde m_t^2} - \frac{|A_t|^4}{6 \tilde m_t^4} \right] ~, \\[6pt]\label{eq:Higgs_spectrum2}
M_{H_2}^2 &\simeq& M_A^2 + \frac{v^2}{2} |\lambda_5| \Big( \cos(\phi_5 + 2\theta) - 1 \Big)  ~, \\[6pt]\label{eq:Higgs_spectrum3}
M_{H_3}^2 &\simeq& M_A^2 + \frac{v^2}{2} |\lambda_5| \Big( \cos(\phi_5 + 2\theta) + 1 \Big)  ~.
\end{eqnarray}
In the second line of~(\ref{eq:Higgs_spectrum}) we also included the dominant 1-loop corrections to the lightest Higgs mass for which they are most relevant. As one notes from the approximate tree level part of the expression in (\ref{eq:Higgs_spectrum}), the BMSSM effects on the lightest Higgs mass are relevant for not too large values of $\tan\beta$ and of the NP scale $M$ (entering the approximate expression through $\lambda_6$). Their sign depends mainly on the phase of $\lambda_6$, namely the phase of $\omega$. Dependence on the phase of $\lambda_5$ and correspondingly on the phase of $\alpha\omega$ arises first at the order $1/(\tan^2\beta \, M)$, i.e. it is only relevant for very small values of $\tan\beta$. This feature is also illustrated in Fig.~\ref{fig:massh} where we compare the value of the mass of the lightest Higgs boson in the $\rm{Arg}(\alpha)-\rm{Arg}(\omega)$ plane, for $\tan\beta = 2$ (left) and $\tan\beta = 5$ (right), fixing the remaining free parameters of the model to a reference point.

In Fig.~\ref{fig:massh}, as well as in the remainder of this work, in the numerical computation of the Higgs spectrum, we always include the 2-loop supersymmetric corrections to the Higgs potential as given in~\cite{Carena:1995bx,Pilaftsis:1999qt} and diagonalize the $3\times 3$ Higgs mass matrix~(\ref{eq:Higgs_mass_matrix}) numerically.

The two plots in Fig.~\ref{fig:massh} show the decoupling of the NP effects on the lightest Higgs mass with $\tan\beta$. For $\tan\beta \gtrsim 10$, the lightest Higgs mass differs from the MSSM expectation only by few~GeV. From the figure, it is also evident that the maximal values for the lightest Higgs mass are obtained in the CP conserving case $\rm{Arg}(\alpha)=\rm{Arg}(\omega)=0$.

\begin{figure}[H] \centering
\includegraphics[width=0.45\textwidth]{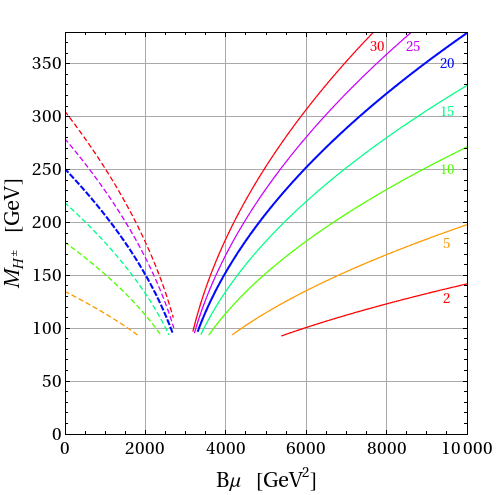} \\[8pt]
\includegraphics[width=0.45\textwidth]{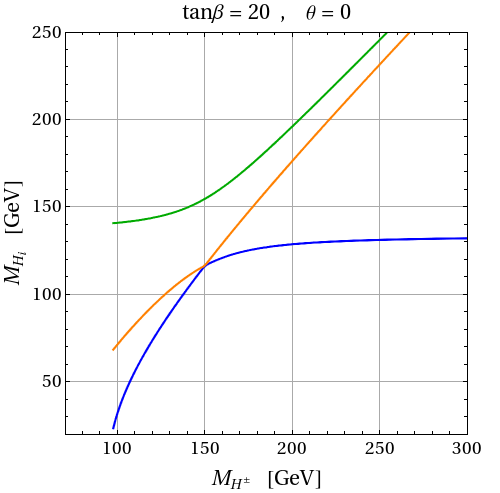} ~~~
\includegraphics[width=0.45\textwidth]{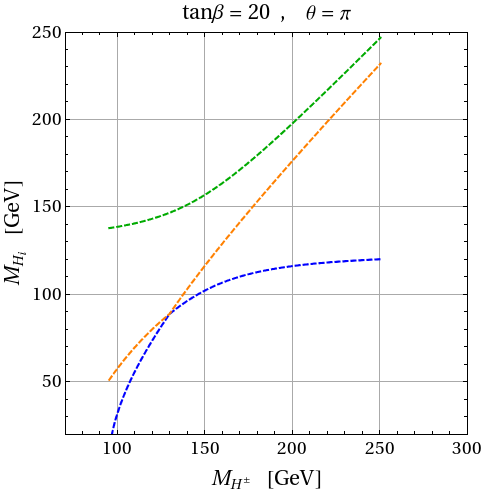} 
\caption{\small 
Top: The charged Higgs mass $M_{H^\pm}$ as a function of the $B\mu$ parameter for different values of $\tan\beta$ as indicated in the plot. Bottom: the three neutral Higgs masses $M_{H_i}$ as a function of $M_{H^\pm}$ for $\tan\beta = 20$. In all plots the remaining parameters are fixed as: $|\alpha| = |\omega| = 1$, $\mu = m_S = 150$~GeV, $M = 1.5$~TeV, $m_{\tilde t} = 800$~GeV, $A_t = 2 m_{\tilde t}$, Arg$(\alpha)=$Arg$(\omega)=0$.
}
\label{fig:double_solution}
\end{figure}

For completeness, we also give the analytical expression for the charged Higgs mass
\begin{eqnarray}\nonumber
M_{H^\pm}^2 &\simeq& M_A^2 + M_W^2 + \frac{v^2}{2} |\lambda_5| \cos(\phi_5 + 2\theta) \\\label{eq:masscharged}
&\simeq& \frac{1}{c_\beta s_\beta} B\mu \cos\theta - \frac{1}{c_\beta s_\beta} \frac{v^2}{2} |\lambda_6| \cos(\phi_6 + \theta) - \frac{v^2}{2} |\lambda_5| \cos(\phi_5 + 2\theta) + M_W^2 ~.
\end{eqnarray}
In presence of CP violation it is customary to characterize the Higgs sector in terms of $\tan\beta$ and $M_{H^\pm}$ instead of $\tan\beta$ and $M_A$, given that $M_A$ is no longer a physical mass.
Unlike in the MSSM however, we note that in the BMSSM there is not necessarily a one to one correspondence between $B\mu$ and $M_{H^\pm}$, already in the CP conserving case.
In the upper plot of Fig.~\ref{fig:double_solution} we show in an example of a CP conserving scenario the charged Higgs mass as a function of $B\mu$ for several values of $\tan\beta$. Choosing for example $\tan\beta = 20$, we observe that charged Higgs masses between $100$~GeV and $250$~GeV can be realized by two different choices of $B\mu$. The non-monotonic dependence of $M_{H^\pm}$ on $B\mu$ arises because the phase of the Higgs VEV that enters Eq. (\ref{eq:masscharged}) changes by varying $B\mu$. In fact, in the CP conserving case, for large enough values of $B\mu$, the minimization condition (\ref{eq:mintheta}) implies $\theta = 0$ and the charged Higgs mass decreases with decreasing $B\mu$ (see solid curves in Fig.~\ref{fig:double_solution}). On the other hand, for very small values of $B\mu$ and a positive $\lambda_6$, the condition (\ref{eq:mintheta}) implies $\theta = \pi$ and the charged Higgs mass increases again for further decreasing $B\mu$ (see dashed curves in Fig.~\ref{fig:double_solution}).\footnote{For intermediate $B\mu$ there exists a region of spontaneous CP violation where the phase of the Higgs VEV changes continuously from $0$ to $\pi$. That region is however not phenomenologically viable as the neutral Higgs spectrum becomes tachionic.}
The range of charged Higgs masses that can be realized by two different values of $B\mu$ becomes smaller for smaller $\tan\beta$ and $|\lambda_6|$ and eventually vanishes. 

Given the fact that one Higgs mass can potentially be realized by two different values of $B\mu$, we conclude that fixing the charged Higgs mass, $\tan\beta$ as well as $\lambda_5$ and $\lambda_6$ does not uniquely determine the Higgs sector of the theory. This is further illustrated in the lower plots of Fig.~\ref{fig:double_solution} that show the neutral Higgs spectrum as function of the charged Higgs mass with all other parameters fixed. The solid curves correspond to large values of the Lagrangian parameter $B\mu$ (and consequently $\theta=0$), while the dashed curves correspond to small $B\mu$ (and consequently $\theta = \pi$). In particular for the lightest Higgs mass one observes a significant shift between the two cases. 

In the scenarios that we are analyzing in the remainder of this work, $\tan\beta$ is fixed to a rather small value and we checked that the given charged Higgs masses do fix the Higgs sector of the model and specifying the $B\mu$ term is not necessary.

\section{Constraints from Electric Dipole Moments}\label{sec:EDMs}

In this section we discuss the constraints coming from Electric Dipole Moments (EDMs) on the new CP violating phases arising in the Higgs sector.

EDMs are known to be highly sensitive probes of CP violation in extensions of the Standard Model~\cite{Pospelov:2005pr}. Indeed the current bounds on experimentally accessible EDMs like the ones of Thallium, Mercury and the neutron~\cite{Regan:2002ta,Griffith:2009zz,Baker:2006ts},
\begin{eqnarray} \label{eq:dTl_exp}
d_{\rm Tl} &\leq& 9.4 \times 10^{-25} ~e\,{\rm cm}~~~@~90\%~ \textnormal{C.L.}~, \\ \label{eq:dHg_exp}
d_{\rm Hg} &\leq& 3.1 \times 10^{-29} ~e\,{\rm cm}~~~@~95\%~ \textnormal{C.L.}~, \\ \label{eq:dn_exp}
d_{n}      &\leq& 2.9 \times 10^{-26} ~e\,{\rm cm}~~~@~90\%~ \textnormal{C.L.}~,
\end{eqnarray}
generically lead to very tight constraints on new sources of CP violation that can be present in extensions of the Standard Model. In particular in the MSSM with SUSY particles at the TeV scale, the flavor diagonal CP violating phases of e.g. the gaugino masses, the higgsino mass parameter and the trilinear couplings are strongly constrained~\cite{Pokorski:1999hz,Ellis:2008zy,Li:2010ax}.

\begin{figure}[t] \centering
\includegraphics[width=0.9\textwidth]{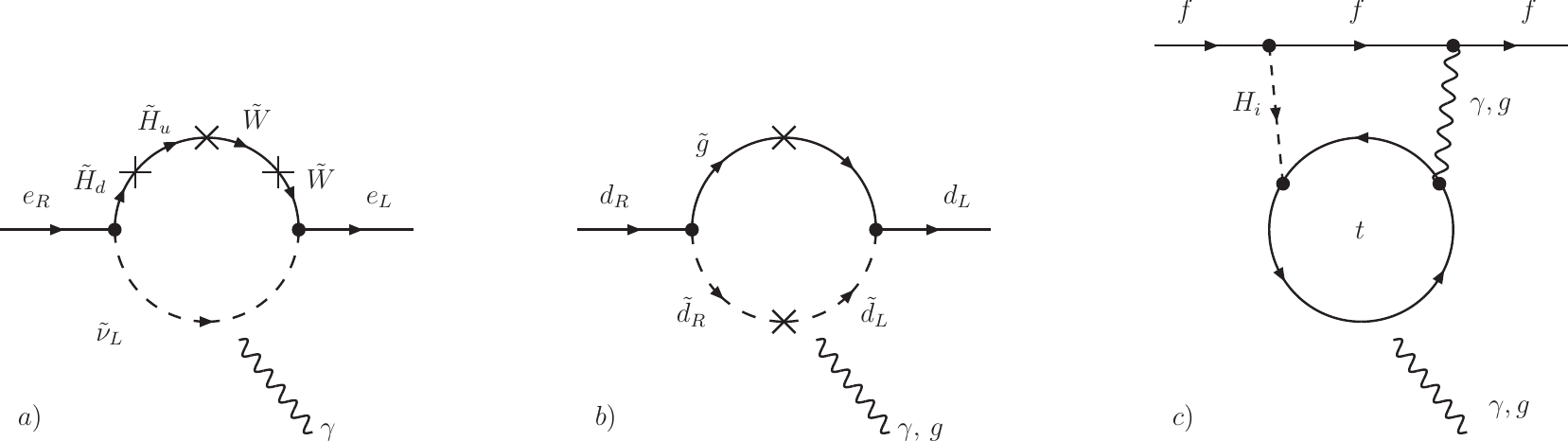}
\caption{\small
Contributions to the electric and chromoelectric dipole moments most relevant for our analysis. Diagram a) is a 1-loop Wino-Higgsino-sneutrino contribution to the electron EDM, while diagram b) is a gluino-down squark contribution to the down quark (C)EDM. Diagram c) represents the dominant 2-loop Barr-Zee type contribution to both the electron EDM and down quark (C)EDM.
}
\label{fig:EDM_diagrams}
\end{figure}

In the following, we take all MSSM parameters real and concentrate on effects of the phases of the BMSSM parameters $\alpha$ and $\omega$.
In this setup important NP effects on the experimentally accessible EDMs are induced by contributions to the lepton and quark (chromo)electric dipole moments ((C)EDMs) both at the 1-loop and at the 2-loop level.\footnote{We explicitly checked that in the scenarios that we consider in the present work, four fermion operators~\cite{Demir:2003js} are always sub-dominant. Still, they are consistently included in our numerical analysis.}
Concerning the 1-loop contributions, note that the higher dimensional operators do not only modify the Higgs sector of the model, but, after electro-weak symmetry breaking, also lead to complex entries in the chargino, neutralino and squark mass matrices (see appendix~\ref{app:SUSY_masses}). In particular, the most important effect arises from the phase of the Higgs VEV that enters these mass matrices, resulting in potentially large 1-loop contributions to both the electron and quark (C)EDMs even if the gaugino and Higgsino masses as well as the trilinear couplings are assumed to be real. In our numerical analysis we include the full set of SUSY 1-loop contributions calculated in the mass eigenstate basis following~\cite{Pokorski:1999hz}. The most important diagrams are a Higgsino-Wino-sneutrino loop for the electron EDM and a gluino-down squark loop for the down quark (C)EDM. They are shown in Fig.~\ref{fig:EDM_diagrams} and have the following approximate expressions
\begin{eqnarray}
\Big\{ d_d^{\tilde g}/e ~,~ \tilde d_d^{\tilde g} \Big\} &\simeq& \frac{\alpha_s}{4\pi} ~m_d~  \textnormal{Im}\left[ e^{i \theta} ~ \frac{t_\beta}{1 + \epsilon_d t_\beta} \right]  ~ \frac{\mu M_{\tilde g}}{\tilde m^4}~ \Big\{ f_d(x_g) ~,~ \tilde f_d(x_g) \Big\} ~, \\[10pt] \label{eq:de_1loop}
d_e^{\tilde H}/e &\simeq& \frac{\alpha_2}{4\pi} ~m_e~ \textnormal{Im}\left[ e^{i \theta} ~ \frac{t_\beta}{1 + \epsilon_\ell t_\beta} \right] ~ \frac{\mu M_2}{\tilde m^4}~ f_e(x_\mu,x_2) ~.
\end{eqnarray}
These expressions show clearly that the 1-loop EDMs are mainly induced by the Higgs phase $\theta$.
The loop functions $f_d$, $\tilde f_d$ and $f_e$ depend on the mass ratios $x_g = M_{\tilde g}^2/\tilde m^2$, $x_\mu = \mu^2/\tilde m^2$ and  $x_2 = M_2^2/\tilde m^2$ and their analytical expressions are given in appendix~\ref{app:loop}. For simplicity we set all sfermion masses to a common mass scale $\tilde m$ in the approximate expressions we show in this work.  
The $\epsilon$ terms appearing in the all order $\tan\beta$ resummation factors, arise from non-holomorphic corrections to the down quark and electron Yukawa couplings. Such corrections become relevant in the large $\tan\beta$ regime and read
\begin{eqnarray}\label{eq:epsilon}
&& \epsilon_d \simeq \epsilon^{\tilde g} + \epsilon^{\tilde H} + \epsilon^{\tilde W}~,~~~ \epsilon_\ell \simeq \epsilon^{\tilde W} ~, \nonumber \\[10pt]
&& \epsilon^{\tilde g} \simeq \frac{\alpha_s}{4 \pi} \frac{8}{3} \frac{\mu M_{\tilde g}}{\tilde m^2} e^{-i \theta} f_1(x_g) ~,~~~
\epsilon^{\tilde H} \simeq \frac{\alpha_2}{4 \pi} \frac{m_t^2}{2 M_W^2} \frac{\mu A_t}{\tilde m^2} e^{-i \theta} f_1(x_\mu)~, \nonumber \\
&& \epsilon^{\tilde W} \simeq - \frac{\alpha_2}{4 \pi}  \frac{3}{2} \frac{\mu M_2}{\tilde m^2} e^{-i \theta} f_2(x_2,x_\mu)~.
\end{eqnarray}
Here we only included gluino, Higgsino and Wino loops, but neglected Bino loops that are typically not relevant. The loop functions $f_1$ and $f_2$ can again be found in appendix~\ref{app:loop}.

It is interesting to note that the 1-loop contributions that are sensitive to the phase of the Higgs VEV are those that involve non-holomorphic couplings of the electron and down quark to the up-type Higgs and are therefore $\tan\beta$ enhanced.\footnote{One loop contributions to the up quark (C)EDM that are sensitive to the phase of the Higgs VEV are thus $1/\tan\beta$ suppressed and only relevant for a very small $\tan\beta \simeq 1$.}
Similarly, also in the expressions of the $\tan\beta$ resummation factors the phase of the Higgs VEV appears.

\bigskip\noindent
At the 2-loop level the most important contributions come from Barr-Zee diagrams, including a top quark loop, that are directly sensitive to the mixing of the scalar and pseudoscalar Higgs bosons~\cite{Pilaftsis:2002fe}. The diagrams are again shown in Fig.~\ref{fig:EDM_diagrams}. Their dominant contribution is $\tan\beta$ enhanced and can be approximated by the following expressions
\begin{eqnarray} \label{eq:de_two loop}
d_e^{(2)t}/e &\simeq& \frac{\alpha_2 \alpha_{em}}{16 \pi^2} \frac{4}{3}  m_e ~  \textnormal{Re}\left[ \frac{\tan\beta}{1 + \epsilon_\ell t_\beta} \right] \frac{m_t^2}{M_W^2} \nonumber \\ 
&& \times \sum_{i=1}^3 \frac{1}{M_{H_i}^2} O_{3i} \left( \frac{s_\alpha}{s_\beta} O_{2i} + \frac{c_\alpha}{s_\beta} O_{1i} \right) f\left(\frac{m_t^2}{M_{H_i}^2}\right) ~,
\end{eqnarray}
\begin{equation} \label{eq:dd_two loop}
\Big\{ d_d^{(2)t}/e ~,~ \tilde d_d^{(2)t} \Big\} \simeq \frac{m_d}{m_e} ~ \frac{\textnormal{Re}[1+ \epsilon_d t_\beta]}{\textnormal{Re}[1 + \epsilon_\ell t_\beta]}  ~ d_e^{(2)t}/e ~ \left\{ \frac{1}{3} ~,~ \frac{\alpha_s}{\alpha_{em}} \frac{3}{8} \right\} ~.
\end{equation}
Here, $O_{ij}$ are the elements of the matrix $O$ defined in~(\ref{eq:O}) that diagonalizes the neutral Higgs mass matrix.
Subleading contributions that are not enhanced by $\tan\beta$ can become important for small values of $\tan\beta$.
The 2-loop function $f$ that enters Eq.~(\ref{eq:de_two loop}) can be found in appendix~\ref{app:loop}. The expression (\ref{eq:de_two loop}) can be further expanded in the decoupling limit and performing an expansion in $1/M$. We find 
\begin{eqnarray}\label{eq:barrzeeexpansion}
&& -\sum_{i=1}^3 \frac{1}{M_{H_i}^2} O_{3i} \left( \frac{s_\alpha}{s_\beta} O_{2i} + \frac{c_\alpha}{s_\beta} O_{1i} \right) f\left(\frac{m_t^2}{M_{H_i}^2}\right) \simeq \nonumber \\
&\simeq&\left( \frac{v^2 |\lambda_5| \sin(\phi_5 + 2\theta)}{2 M_A^4} - \frac{v^4 |\lambda_6|^2 \sin(2\phi_6 + 2\theta)}{2 M_A^4 M_h^2} \right) \left[ f\left(z\right) + z ~ \partial_z f\left(z\right) \right]~,
\end{eqnarray}
with the mass ratio $z = m_t^2/M_A^2$. Eq.~(\ref{eq:barrzeeexpansion}) clearly shows that the 2-loop EDMs are directly induced by the new phases of the higher dimensional operators.
In the scenarios we consider, the Higgs bosons are always light compared to sfermions and the 2-loop contributions can compete with or even dominate the 1-loop contributions discussed above. 
If the stop mass is small, then in addition also 2-loop diagrams with stop loops are often relevant. For large values of $\tan\beta$, (s)bottom and (s)tau loops can also become important. 
Even though the sparticle masses are rather large in the scenarios that we consider in the following, we include the full set of 2-loop Barr-Zee contributions from~\cite{Chang:1998uc,Pilaftsis:2002fe,Demir:2003js} in our numerical analysis.

\bigskip\noindent
Expressing the experimentally accessible EDMs of Thallium, Mercury and the neutron through the quark and electron (C)EDMs induces sizable uncertainties related to QCD, nuclear and atomic interactions. Approximately one finds the following relations~\cite{Pospelov:2005pr,Raidal:2008jk}
\begin{eqnarray} \label{eq:dTl}
d_{\rm Tl} &\simeq& -585 d_e ~, \\ \label{eq:dHg}
d_{\rm Hg} &\simeq&  7 \times 10^{-3} e (\tilde d_u - \tilde d_d) + 10^{-2} d_e~, \\ \label{eq:dn}
d_n &\simeq& 1.4 (d_d - 0.25 d_u) + 1.1 e (\tilde d_d + 0.5 \tilde d_u)~.
\end{eqnarray}
The quark (C)EDMs in the above expressions are understood to be evaluated at a scale of 1~GeV. Expressions for the running of the EDMs down from the high matching scale can be found e.g. in~\cite{Barbieri:2011vn}.
While the prediction for the Thallium EDM is rather robust, the uncertainty in the neutron EDM is estimated to be at the level of 50\% and the expression for the Mercury EDM is only accurate up to a factor of 2-3~\cite{Raidal:2008jk}. We take these uncertainties into account when evaluating the corresponding constraints.

\begin{figure}[t] \centering
\includegraphics[width=0.45\textwidth]{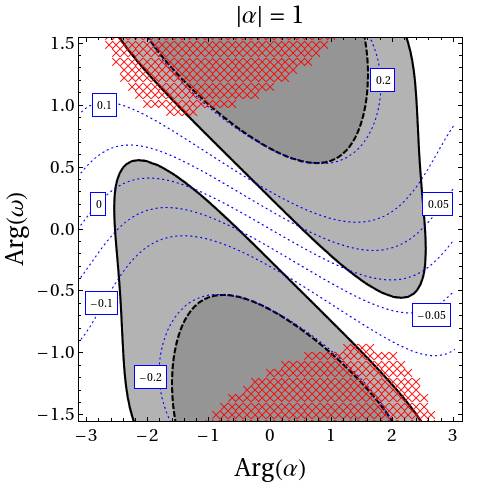} ~~~
\includegraphics[width=0.45\textwidth]{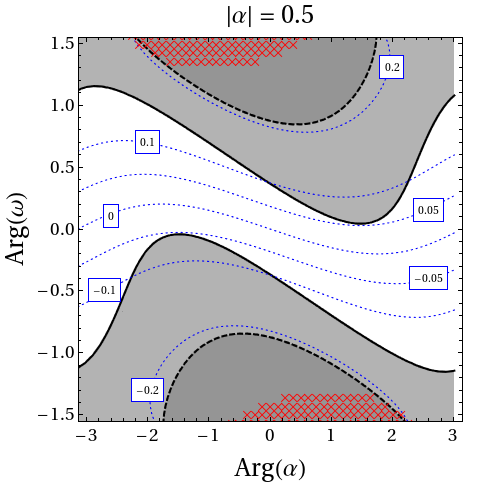} 
\caption{\small
Bounds from the EDMs in the Arg$(\alpha)$ - Arg$(\omega)$ plane for two different values of $|\alpha| = 1, 0.5$. The remaining parameters are fixed to $\tan\beta = 2$, $|\omega| = 1$, $\mu = m_S = 150$~GeV, $M = 1.5$~TeV, $M_{H^\pm} = 200$~GeV, $\tilde m = 800$~GeV, $A_t = 2 \tilde m$, $A_b = A_\tau = 0$ and $M_{\tilde g} = 3 M_2 = 6 M_1 = 1200$~GeV. The solid and dashed black lines correspond to the Thallium and Mercury EDMs respectively. The neutron EDM does not lead to constraints in the considered scenarios. The dotted blue lines indicate the values of the phase of the Higgs VEV $\theta$. In the red hatched region the electro-weak symmetry breaking vacuum is only a local minimum of the Higgs potential.
}
\label{fig:EDM_constraints}
\end{figure}

As the dominant contributions to the EDMs are $\tan\beta$ enhanced, larger values of $\tan\beta$ lead to stronger constraints. Here we restrict the discussion to the low $\tan\beta$ regime, where we expect the most interesting Higgs collider phenomenology (see Sec.~\ref{sec:collider_pheno}). A detailed treatment of EDMs for large $\tan\beta$ will be important when we explore a complementary region of parameter space analyzing interesting effects in the flavor phenomenology (see Sec.~\ref{sec:Bsmixing}).

In Fig.~\ref{fig:EDM_constraints} we show the constraints coming from the EDMs in the $\rm{Arg}(\omega)-\rm{Arg}(\alpha)$ plane for two example scenarios with $\tan\beta = 2$. We observe that the most stringent constraint comes from the Thallium EDM that is dominantly induced by the 1-loop chargino contribution to the electron EDM. From (\ref{eq:de_1loop}) one expects that the allowed region therefore corresponds to small values of the phase of the Higgs VEV $\theta$ as it is also shown in Fig.~\ref{fig:EDM_constraints}, where the values of $\theta$ are indicated by the dotted blue contours.
In~(\ref{eq:de_1loop}) we neglected additional $1/M$ suppressed corrections that can be incorporated by replacing the Higgsino mass with an effective term $\mu e^{i\theta} \to \mu e^{i\theta} - \omega \frac{v^2}{M} s_\beta c_\beta e^{2i\theta}$.
Therefore, the region compatible with the Thallium EDM is slightly tilted with respect to the $\theta = 0$ line in the plots of Fig.~\ref{fig:EDM_constraints}.

Concerning the Mercury EDM, we find that generically it is mainly induced by the 2-loop CEDM of the down quark in Eq.~(\ref{eq:dd_two loop}). Interestingly, also the regions allowed by the Mercury EDM correspond to small values of the phase of the Higgs VEV in Fig.~\ref{fig:EDM_constraints}. Given its large theory uncertainty however, the Mercury EDM is typically less constraining than the Thallium EDM.

It is instructive to derive from Eq.~(\ref{eq:mintheta}) an analytical expression for the $\theta = 0$ line where the EDM constraints are in general easier satisfied. Defining  $\phi_\omega = \rm{Arg}(\omega)$ and $\phi_\alpha = \rm{Arg}(\alpha)$, one has
\begin{equation}
\sin \phi_\omega = - s_\beta c_\beta \frac{|\alpha| m_S}{\mu} \sin (\phi_\alpha + \phi_\omega) ~.
\end{equation}
This shows that for smaller values of $|\alpha|$ and larger values of $\tan\beta$, the allowed region corresponds to smaller values of $\phi_\omega$. 
Usually one finds that the full range $-\pi < \rm{Arg}(\alpha) < \pi$ remains allowed if Arg$(\omega)$ is chosen appropriately. For this reason, in the following we will study observables in the Higgs collider phenomenology mainly as function of Arg$(\alpha)$.

\section{Higgs Collider Phenomenology} \label{sec:collider_pheno}

\subsection{LEP and Tevatron Bounds}\label{sec:LEPbounds}

In this section we study the LEP and Tevatron bounds on the Higgs bosons of the model.
In order to evaluate the bounds from direct Higgs searches on the BMSSM parameter space, knowledge of the couplings of the Higgs bosons is required. 
It is convenient to work with effective couplings that are normalized to the corresponding SM couplings.
For the effective $H_iZZ$ and $H_iWW$ couplings, that are essential both for the Higgs production at LEP and the Higgs decays at the Tevatron and the LHC, one finds expressions that have the same structure as in the MSSM with CP violation
\begin{eqnarray}
\xi_{ZZH_i} = \xi_{WWH_i} &=& s_{\beta - \alpha} O_{1i} + c_{\beta-\alpha} O_{2i} ~. 
\end{eqnarray}
We stress however that the angle $\alpha$ and the rotation matrix $O$ differ from the MSSM. They are highly sensitive to the higher dimensional operators of the BMSSM and given in Eqs.~(\ref{eq:alpha}),~(\ref{eq:calpha}) and~(\ref{eq:O}).
The couplings obey the relation
\begin{equation}
\sum_i \xi^2_{ZZH_i} = 1~.
\end{equation}
Possible deviations from the above expressions arise from $1/M^2$ suppressed operators~\cite{Carena:2009gx} that we however do not consider here. Similarly, in the considered framework also the Higgs-quark and Higgs-lepton couplings have the same structure as in the MSSM with CP violation, with the angle $\alpha$ and the rotation matrix $O$ given by their BMSSM expressions~(\ref{eq:alpha}),~(\ref{eq:calpha}) and~(\ref{eq:O}).

We calculate the effective $H_igg$ and $H_i\gamma\gamma$ couplings from the ratios of the LO decay widths to the LO SM decay widths
\begin{eqnarray}
\xi^2_{\gamma\gamma H_i} &=& \frac{\Gamma(H_i \to \gamma\gamma)^{\rm LO}}{\Gamma(H_i \to \gamma\gamma)_{\rm SM}^{\rm LO}} ~, \\
\xi^2_{ggH_i} &=& \frac{\Gamma(H_i \to gg)^{\rm LO}}{\Gamma(H_i \to gg)_{\rm SM}^{\rm LO}} \simeq \frac{\sigma(gg \to H_i)_{\rm \phantom{SM}}}{\sigma(gg \to H_i)_{\rm SM}} ~. \label{eq:Hggcoupling}
\end{eqnarray}
In our calculation we incorporate the full set of SM and SUSY particle contributions using the expressions in~\cite{Lee:2003nta}. As stated in Eq.~(\ref{eq:Hggcoupling}), we assume that the effective $H_igg$ coupling approximates the ratio between the $gg \to H_i$ production cross sections at Tevatron and LHC and the corresponding SM cross sections. This approach has also been adopted in~\cite{Carena:2010cs}, where it has been explicitly checked that it leads to results in the CP conserving BMSSM that are accurate within 5\%~-~20\%, depending on $\tan\beta$. We do not expect this to change in presence of CP violation.

To check compatibility with Higgs searches at LEP and Tevatron, we use the latest version of \verb|Higgsbounds|~\cite{Bechtle:2008jh,Bechtle:2011sb} in the effective coupling approximation. 
To obtain the total decay width of the Higgs bosons of the BMSSM, we make use of the results collected in~\cite{Lee:2003nta}, replacing the MSSM couplings with the BMSSM ones.
In the computation of the several partial decay width, we use ratios of partial decay widths in the BMSSM and the SM and multiply the results with the state of the art SM partial decay widths obtained from \verb|HDECAY|~\cite{Djouadi:1997yw}, when applicable.
We also cross checked our results using a version of \verb|CPsuperH|~\cite{Lee:2003nta,Lee:2007gn} with appropriately modified couplings.

We remark that by default, \verb|Higgsbounds| uses the latest combined SM Higgs exclusion from the Tevatron~\cite{Aaltonen:2011gs} only for Higgs bosons that satisfy very restrictive requirements on their ``SM-likeness''. In particular, a Higgs boson is considered SM-like by \verb|Higgsbounds| if its different cross sections normalized to the SM values differ at most by 2\% from a common scale factor. For Higgs bosons that do not satisfy this requirement, the strongest constraint used is then typically the $gg \to H_i \to WW$ analysis in~\cite{Aaltonen:2010sv} that is considerably weaker.
In our BMSSM scenarios discussed below, the most distinct cases are those in which the three neutral Higgs bosons share couplings to gauge bosons and fermions in a non SM-like way.
In order to get a reasonable estimate of the current Tevatron bounds on the BMSSM parameter space, we therefore consider in addition to \verb|Higgsbounds| also the latest Tevatron exclusion. As in the region that is excluded by Tevatron Higgs searches, the dominant process is $gg \to H_i \to W W$, we compute the corresponding cross section in the BMSSM normalized to the SM and apply the bounds given in~\cite{Aaltonen:2011gs} for Higgs masses in the range $145$~GeV $< M_{H_i} < 200$~GeV, where the effect of vector boson fusion and associated production is minimal.

Also in the low mass region, \verb|Higgsbounds| does not use the combined Tevatron exclusion~\cite{:2010ar}, but applies separately the different analyses entering the combination, for most of the BMSSM parameter space. We find however that even the combined low mass bounds from~\cite{:2010ar}, that are dominated by search channels where the Higgs is produced in association with a vector boson and decays into $b \bar b$, are not strong enough yet to exclude BMSSM parameter space.

\begin{figure}[t] \centering
\includegraphics[width=0.51\textwidth]{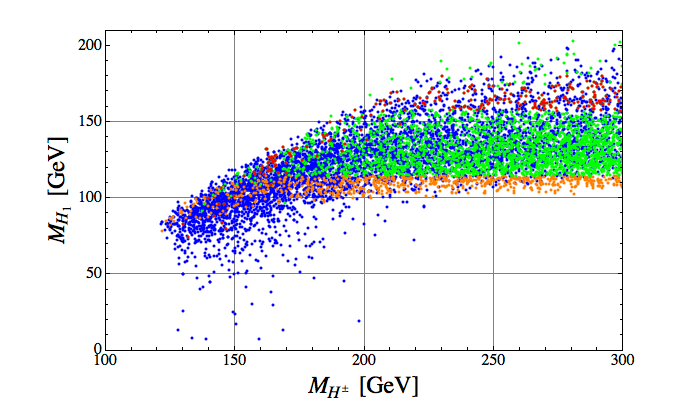} ~
\includegraphics[width=0.45\textwidth]{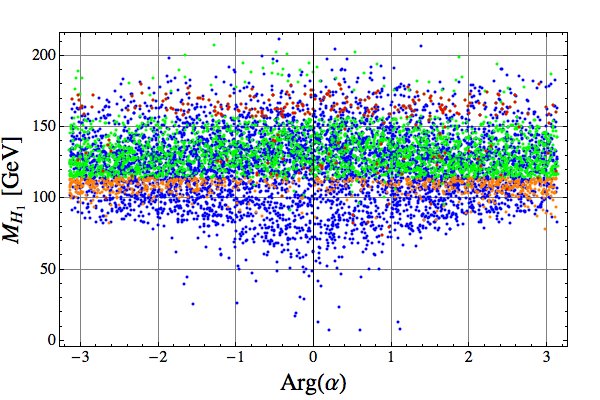}
\caption{\small 
Lightest Higgs boson mass as a function of the charged Higgs boson mass (left) and the phase of $\alpha$ (right). Shown are the points excluded by EDM constraints (in blue), LEP (in orange) and Tevatron (in red). Green points are allowed. The scan is performed fixing $\tan\beta=2$, $m_S=\mu=150$~GeV, a common squark mass of $\tilde m = 800$~GeV, a common slepton mass of $\tilde m_\ell = 1100$~GeV, trilinear couplings $A_t = 2 \tilde m$, $A_b = A_\tau = 0$ and gaugino masses $M_{\tilde g} = 3M_2 = 6M_1 = 1200$~GeV and scanning over $\alpha,\,\omega,\,M,\, M_{H^\pm}$ in the ranges $\alpha = (0.5-2)e^{i\rm{Arg}(\alpha)}$ with Arg$(\alpha)\in[0,2\pi]$, $\omega = (0.5-2)e^{-\frac{i}{5}\rm{Arg}(\alpha)}$, $M = (1-3)$~TeV and $M_{H^\pm}<350$~GeV, respectively.}
\label{fig:scan}
\end{figure}

In Fig.~\ref{fig:scan} we present the result of a parameter scan of the model as defined in the figure caption, for the mass of the lightest Higgs boson as a function of the mass of the charged Higgs on the left and as a function of the phase of $\alpha$ on the right. All the shown points satisfy the requirement of a correct EWSB (see Sec.~\ref{sec:vacuum_stability}).
In addition, we impose constraints from EDMs (points in blue are excluded), then we check the compatibility with LEP using \verb|Higgsbounds| (points in orange are excluded) and finally we impose the Tevatron bounds as described previously (points in red are excluded). It is interesting to note that points satisfying all the constraints (in green/light gray) arise in the entire range for the phase of $\alpha$, i.e. in spite of the strong constraints coming from EDMs, large CP violating phases are allowed in the model.

As expected, due to the higher dimensional operators, the values of the lightest Higgs mass cover a much larger range as compared to the MSSM and reach up to $\simeq 210$~GeV, even in presence of large CP violating phases (see right panel of Fig.~\ref{fig:scan}).
As discussed in~\cite{Carena:2009gx} in the CP conserving case, effects of higher dimensional operators at the $1/M^2$ order can increase that limit further up to $\simeq 300$~GeV. As we will discuss in the next section, a rather heavy lightest Higgs boson (in the range (170-210)~GeV) has usually a gluon gluon fusion production cross section times branching ratio into vector bosons that is enhanced with respect to the SM. Such a Higgs boson will be easily probed at the LHC already with 1~fb$^{-1}$ (see Fig.~\ref{fig:collider_sensitivity} below). 

The few allowed points below the SM LEP bound of 114.4~GeV, correspond to relatively light charged Higgs masses ($\sim (130-170)$~GeV). These points can be reached both in the CP conserving and in the CP violating cases (see right panel of Fig.~\ref{fig:scan}) and arise once the lightest Higgs boson is close to a pseudo scalar state, namely with a strongly reduced coupling to gauge bosons. Correspondingly, such Higgs bosons are very difficult to probe both at the Tevatron and at the LHC.

\subsection{Characteristic Scenarios for Collider Searches} \label{sec:scenarios}

In this section we present our analysis of the Higgs collider phenomenology of the model. Taking into account constraints from vacuum stability and EDMs as well as collider constraints from LEP and the Tevatron we concentrate on those aspects that can distinguish the BMSSM with CP violation from the case without CP violation that has been extensively studied in~\cite{Carena:2010cs}.

Genuine signatures of the new CP violating phases of the BMSSM can in principle occur in CP asymmetries based on the longitudinal $\tau$ polarization in the $WW \to H_i \to \tau \tau$ channel. Such asymmetries can be resonantly enhanced in scenarios where two (or all three) neutral Higgs bosons are nearly degenerate, with mass differences comparable to their decay widths~\cite{Ellis:2004fs}.
A measurement of these observables however appears to be challenging at the LHC and a detailed study of them is beyond the scope of the present work. We concentrate instead on distinct features in the Higgs spectrum, the Higgs - vector boson couplings, the $gg \to H_i$ production cross sections and the Higgs branching fractions that we analyze first in a generic scan of the parameter space (Sec.~\ref{sec:scangen}) and secondly in several representative scenarios (Sec.~\ref{sec:3scenarios}). We calculate the Higgs branching ratios using the results collected in~\cite{Lee:2003nta} in combination with \verb|HDECAY|~\cite{Djouadi:1997yw}, after implementing the BMSSM Higgs spectrum and couplings appropriately.

As the higher dimensional operators of the BMSSM have the strongest impact on the mass of the lightest neutral Higgs boson for small values of $\tan\beta$, we concentrate our discussion here to scenarios in the low $\tan\beta$ regime. This has the additional advantage that constraints from EDMs are kept at a minimum.

\subsubsection{Generic Features of the Parameter Scan}\label{sec:scangen}

In the mass range $M_{H_i} \gtrsim 140 (120)$~GeV, the process $gg \to H_i \to WW/ZZ$ is the dominant one for Higgs searches at the Tevatron (LHC). For lower Higgs masses, $M_{H_i} < 125$~GeV, the associated production with subsequent decay into $b \bar b$, $V \to V H_i \to b \bar b$, is the main search channel at the Tevatron. At the LHC on the other hand, the $gg \to H_i \to \gamma\gamma$ channel is the most important one for low Higgs masses. The production in vector boson fusion and decay into $\tau^+ \tau^-$, $WW \to H_i \to \tau^+ \tau^-$, as well as the associated production of boosted Higgs bosons that decay into $b \bar b$, are only marginally important at the 7~TeV LHC run. We remark that the BMSSM effects in the Higgs couplings can of course modify the relative importance of the different search channels.
In the following we consider a parameter scan, analyzing the most relevant Higgs search channels mentioned above.
We use the same parameter scan already described in the previous section for the study of LEP and Tevatron bounds.

\begin{figure}[H] \centering
\includegraphics[width=0.37\textwidth]{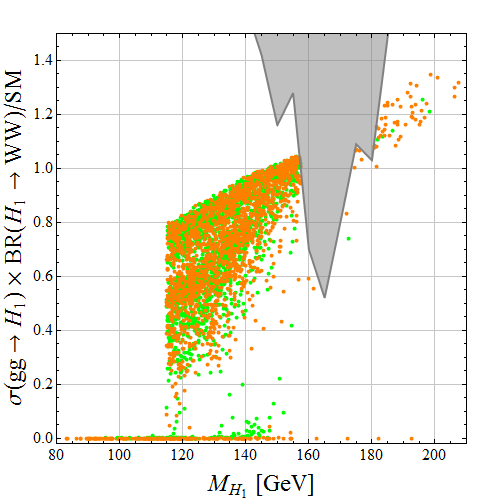} ~~~
\includegraphics[width=0.37\textwidth]{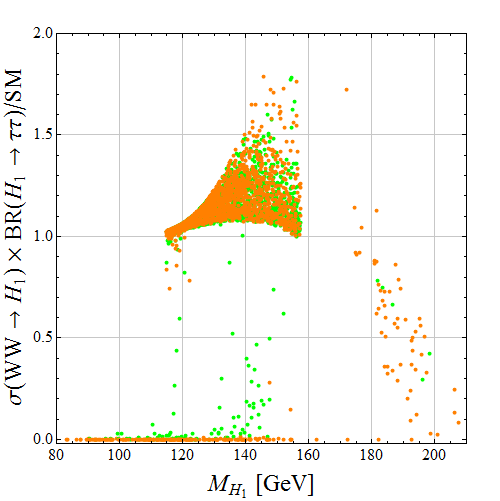} \\
\includegraphics[width=0.37\textwidth]{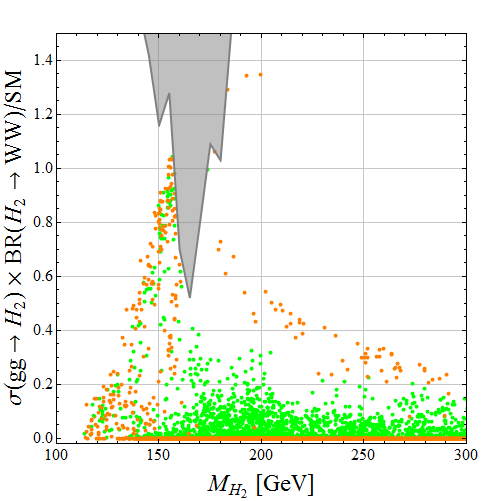} ~~~
\includegraphics[width=0.37\textwidth]{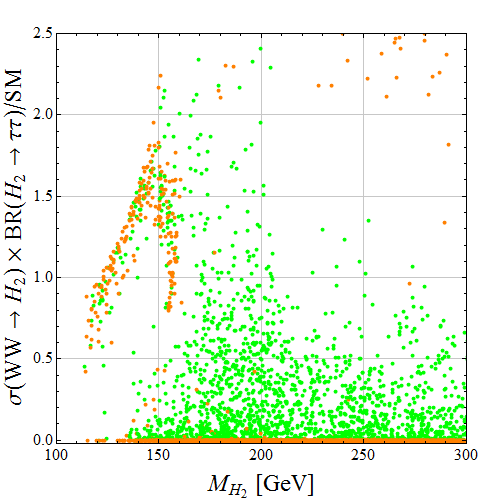} \\
\includegraphics[width=0.37\textwidth]{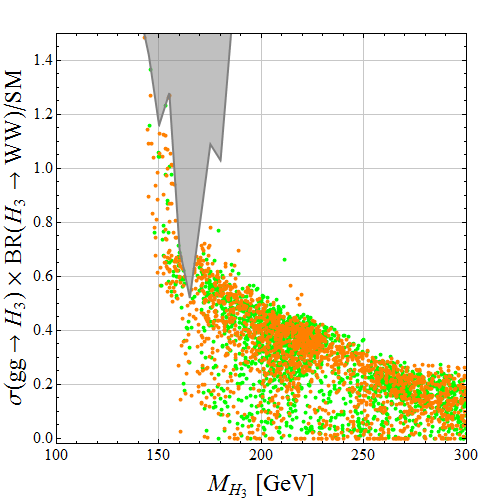} ~~~
\includegraphics[width=0.37\textwidth]{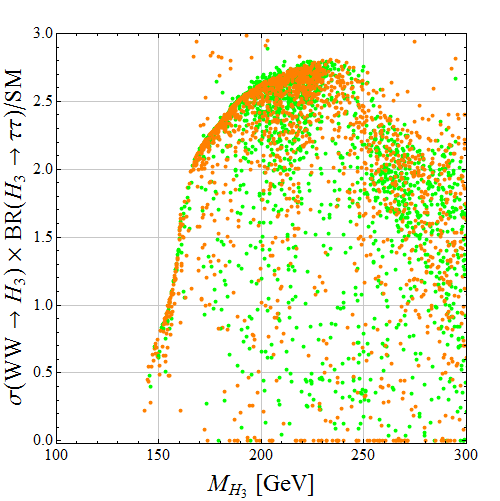}
\caption{\small 
Left: $gg \to H_i$ production cross sections times Higgs branching ratios into $WW$ normalized to the SM values. Right: $WW \to H_i$ production cross sections and Higgs branching ratios into $\tau\tau$ normalized to the SM values. Orange (dark gray) points correspond to approximate CP conservation ($|$Arg$(\alpha,\omega)| < 0.1$); green (light gray) points correspond to the CP violating case ($|$Arg$(\alpha,\omega)| > 0.1$).
Only points allowed by LEP and Tevatron bounds, vacuum stability and EDMs are shown. The gray region is excluded by the latest combined Tevatron analysis with 8.2~fb$^{-1}$~\cite{Aaltonen:2011gs}. See caption of Fig.~\ref{fig:scan} for the details of the scan.}
\label{fig:scan2}
\end{figure}

The left panels of Fig.~\ref{fig:scan2} show the product of the $gg \to H_i$ production cross sections and Higgs branching ratios into $WW$, normalized to the SM values. In the right panels instead, the cross section of the complementary LHC channel $WW \to H_i \to \tau\tau$ normalized to the SM cross section is shown. 
For the low values of $\tan\beta$ considered here, the effective $H_i b\bar b$ and $H_i \tau \bar\tau$ couplings are the same to an excellent approximation. In addition, also the ratios of the Higgs production cross sections in vector boson fusion and associated production with the corresponding SM values are identical. Correspondingly, the $WW \to H_i \to \tau\tau$ plots are also valid for the low mass Higgs search at the Tevatron in the $V \to V H_i \to V b\bar b$ channel and similarly also for the boosted Higgs analysis at LHC~\cite{Butterworth:2008iy,Atlasboosted}.
In Fig.~\ref{fig:scan2}, we do not present instead the cross sections of the $gg \to H_i \to \gamma\gamma$ channel, since we find in the case of CP violation, that the lightest Higgs boson generically does not show any sizable enhancement compared to the SM predictions (see also Sec.~\ref{sec:diphoton} below). The two heavier Higgs bosons show a considerable enhancement only for rather heavy masses ($\gtrsim 250$~GeV) for which the di-photon branching ratio is tiny anyway.

From the first row of Fig.~\ref{fig:scan2}, we observe that typically there are good prospects for the detection of the lightest Higgs boson in the $WW$ channel, both in the CP conserving case (orange/dark gray) and in the CP violating case (green/light gray). For $M_{H_1}\lesssim 140$~GeV, the product $\sigma\left(gg\to H_1\right)$ ${\rm BR}\left(H_1\to WW\right)$ is suppressed with respect to the SM; still for many points of the scan, the suppression is not large (in the range 0.8-0.6) and does not prevent the possibility to probe the lightest Higgs boson in the $WW$ channel at the Tevatron and the LHC. We also note that the upper bound on the product $\sigma\left(gg\to H_1\right){\rm{BR}}\left(H_1\to WW\right)$ is due to the maximal value for the charged Higgs mass that we allow in the scan: taking larger masses for the charged Higgs boson would allow to reach values closer to 1 for smaller $M_{H_1}$. 

Concerning the decay of the lightest Higgs to $\tau\tau$, we note typically a slight enhancement over the SM prediction, even in the low mass region ((115-130)~GeV). This enhancement is connected to a suppression of the $H_i \to WW$ branching ratio as observed above.
Given the rather limited sensitivity of ATLAS and CMS for the $\tau\tau$ channel~\cite{ATLAS,CMS}, the enhancement is however not sufficient to probe the Higgs boson in this channel alone at the ongoing 7~TeV LHC run.

Similarly, also the two heavier Higgs bosons can only be discovered in the $WW$ channel. The $\tau\tau$ channel is in fact sensibly enhanced over the SM prediction only for masses above (140-150)~GeV, for which the (not SM-normalized) $\sigma\left(WW\to H_{2,3}\right){\rm{BR}}\left(H_{2,3}\to \tau\tau\right)$ is too small to allow the detection of the heavier Higgs bosons.

In the case of CP conservation, the second lightest Higgs is typically the pseudoscalar state and therefore does not couple to gauge bosons. In the presence of CP violation however, also the second lightest Higgs has non zero couplings to $WW$ and $ZZ$ and sizable $gg \to H_2 \to WW$ and $WW \to H_2 \to \tau \tau$ cross sections arise generically.
On the other hand, it is interesting to note that the $gg \to H_2 \to WW$ cross section of the second lightest Higgs can even be larger than the SM prediction both in the CP violating case and even in the CP conserving case. This happens in regions of parameter space where the pseudoscalar is either the lightest or the heaviest Higgs.  

In general we observe, that from the scatter plots shown in Fig.~\ref{fig:scan2} it is rather difficult to distinguish the CP conserving from the CP violating case, in particular for the lightest and the heaviest Higgs boson, since many features that are characteristic for CP violation are obscured by the general parameter scan.
In the following we therefore discuss several concrete scenarios that allow a distinction between the CP violating and the CP conserving BMSSM. We analyze scenarios with an enhanced di-photon signal, scenarios with Higgs cascade decays and in particular scenarios with three neutral heavy Higgs bosons that show features that are characteristic for the presence of CP violation.

\subsubsection{CP Conserving BMSSM Scenarios} \label{sec:diphoton}

The BMSSM without CP phases is of course a subset of the parameter space that we are considering here.
In the following we comment on two benchmark scenarios which are present already in the absence of CP phases but become rather fine tuned once CP phases are turned on.

\paragraph{Scenarios with Enhanced Di-Photon Signal:}

As discussed in~\cite{Carena:2010cs} for the framework of the BMSSM without additional sources of CP violation, scenarios with an enhanced di-photon signal are in principle possible only in regions of parameter space where the $H_1 \bar b b$ coupling is strongly suppressed with respect to the SM. In that way enhancements in the $gg \to H_1 \to \gamma\gamma$ cross sections up to an order of magnitude compared to the SM can be achieved. Such scenarios are also present in our framework. However, for non-trivial phases they are much more difficult to realize. Neglecting for simplicity $\tan\beta$ resummation corrections, the $H_1 \bar b b$ coupling in presence of CP violation reads
 \begin{equation}
\mathcal{L} = -\frac{m_d}{v} \tan\beta ~ \Big[ \left( \frac{c_\alpha}{s_\beta} O_{21} - \frac{s_\alpha}{s_\beta} O_{11} \right) \bar b b H_1 - O_{31} \bar b i\gamma_5 b H_1 \Big] ~.
\end{equation}
In the absence of CP violation one has $O_{21} = O_{31} = 0$, $O_{11} = 1$ and a suppressed $H_1 \bar b b$ coupling simply corresponds to $s_\alpha \ll c_\beta$. In the presence of CP violation however one has to simultaneously ensure $O_{31} \ll 1/\tan\beta$ and $c_\alpha O_{21} - s_\alpha O_{11} \ll c_\beta$. This requires a large amount of fine tuning between several parameters. We conclude that a strongly enhanced di-photon signal would not be a characteristic signal of CP phases in the BMSSM, but would be a hint towards the absence of additional sources of CP violation.

\paragraph{Scenarios with Higgs Cascade Decays:}

As discussed in~\cite{Carena:2010cs,Bae:2010cd}, multi-Higgs decay chains can be realized in the BMSSM without CP phases. The scenarios considered in these works contain a (very) light pseudoscalar Higgs that avoids the LEP bound as it does not couple to the Z boson. The higher dimensional operators in the BMSSM ensure that the two scalar Higgs bosons are heavy enough in order not to violate LEP constraints. The $h/H \to AA \to 2b 2\tau$ or $4\tau$ decays can then be the most promising search channels for the scalar Higgs bosons.
However, in the presence of CP violation the three Higgs bosons mix and therefore the would-be pseudoscalar will always couple to some extent to gauge bosons. Therefore, for non-trivial phases it is typically more difficult to have a Higgs boson considerably below 114.4~GeV while avoiding the LEP bounds. 

In addition we stress that in scenarios with a light pseudoscalar-like Higgs boson, it is very difficult to simultaneously fulfill constraints coming from vacuum stability and the lightest chargino mass. As seen in Sec.~\ref{sec:vacuum_stability}, the vacuum stability constraint then becomes even stronger once CP phases are introduced. Furthermore, due to the very light Higgs spectrum also EDM constraints are extremely stringent in such a scenario and allow for non-trivial phases only in very fine tuned strips in parameter space.
Again we conclude that scenarios with Higgs cascade decays into a light pseudoscalar-like Higgs are not characteristic of the CP violating BMSSM but more generic in absence of CP violation.

\subsubsection{Characteristic CP Violating Scenarios with 3 Heavy Higgs Bosons}\label{sec:3scenarios}

One of the distinct features of the BMSSM is the possibility to have three heavy neutral Higgs bosons with masses $M_{H_i} \gtrsim 140$~GeV. In the following we discuss three concrete scenarios in which this is realized. The input parameters for all the scenarios are conveniently collected in Tab.~\ref{tab:scenarios_input}.
A detailed discussion of the Tevatron and LHC sensitivities to the scenarios is given at the end of the section. (For a detailed analysis of the Tevatron sensitivity to MSSM Higgs bosons see \cite{Draper:2009fh,Draper:2009au})

\begin{table}[t] 
\addtolength{\arraycolsep}{10pt}
\renewcommand{\arraystretch}{1.3}
\centering
\begin{tabular}{|c|c|c|c|c|}
\hline\hline
& Sc. Ia & Sc. Ib & Sc. II \\ 
\hline\hline
$|\alpha|$ & $1$ & $1$ & $0.8$ \\ 
$|\omega|$ & $2$ & $1$ & $1.6$\\ 
$\textnormal{Arg}(\alpha)$ & $\pi/2$ & $\pi/4$ & $-2\pi/3$ \\ 
$\textnormal{Arg}(\omega)$ & $-\pi/10$ & $-\pi/20$ & $\pi/20$\\ 
$\tan\beta$ & $2$ & $2$ & $3$ \\ 
$M_{H^\pm}$ [GeV] & $195$ & $225$ & $166$ \\ 
$M$ [TeV]& $2.5$ & $2$ & $2$ \\ 
$\mu$ [GeV]& $160$ & $190$ & $140$ \\ 
$m_S$ [GeV]& $160$ & $400$ & $100$ \\ 
\hline\hline
\end{tabular}
\caption{\small
Input parameters for the scenarios discussed in the present section. For all scenarios we choose a common squark mass of $\tilde m = 800$~GeV, a common slepton mass of $\tilde m_\ell = 1100$~GeV, trilinear couplings $A_t = 2 \tilde m$, $A_b = A_\tau = 0$ and gaugino masses $M_{\tilde g} = 3M_2 = 6M_1 = 1200$~GeV.
}
\label{tab:scenarios_input}
\end{table}

\paragraph{Scenario Ia:}

\begin{figure}[H] \centering
\includegraphics[width=0.39\textwidth]{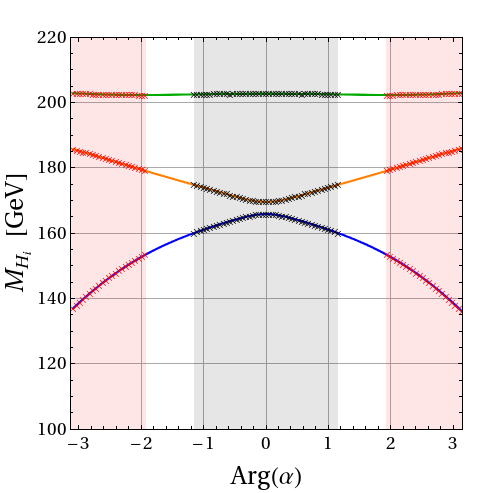} ~~~
\includegraphics[width=0.39\textwidth]{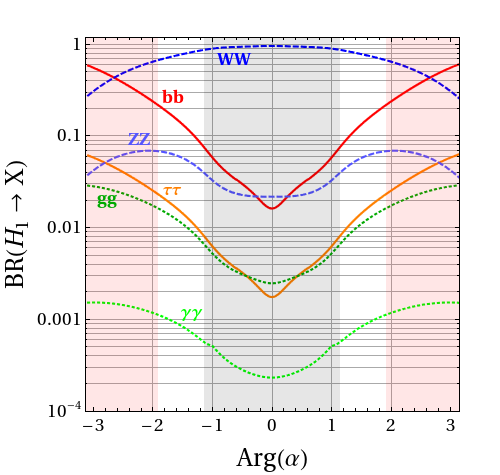} \\
\includegraphics[width=0.39\textwidth]{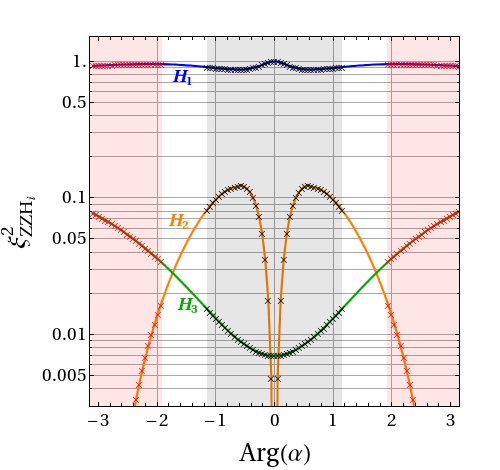} ~~~
\includegraphics[width=0.39\textwidth]{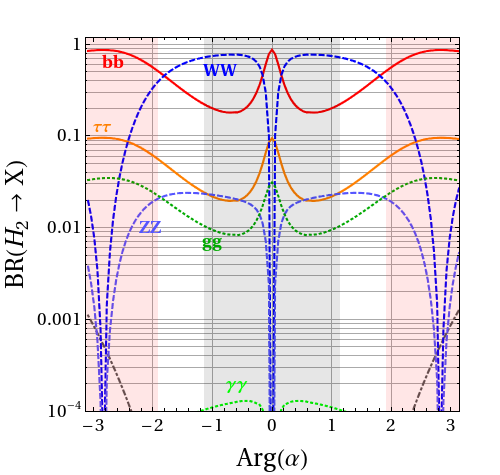} \\
\includegraphics[width=0.39\textwidth]{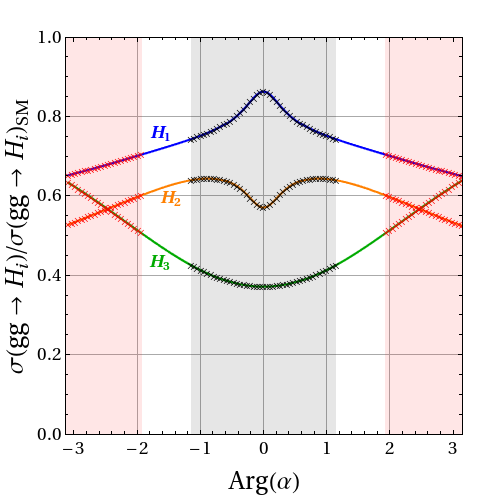} ~~~
\includegraphics[width=0.39\textwidth]{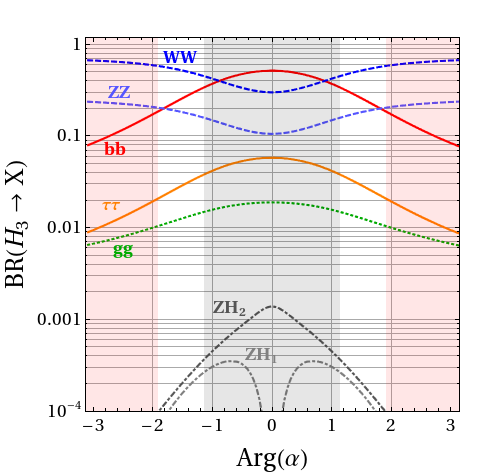}
\caption{\small 
Higgs spectrum, Higgs - vector boson couplings, $gg \to H_i$ production cross sections and Higgs branching ratios as function of the phase of $\alpha$ for the Scenario Ia. The phase of $\omega$ is fixed to Arg$(\omega) = -$Arg$(\alpha)/5$ in the plots, to keep under control EDM constraints. The side bands (light red shaded regions) correspond to vacua that are only local minima of the potential and the central band (light gray shaded region) is excluded by Tevatron Higgs searches.}
\label{fig:sce1}
\end{figure}

In the first scenario, all three neutral Higgs bosons are heavier than 150~GeV and decay dominantly into $WW$. This scenario cannot be realized, either in the BMSSM without CP violating phases or in the CP violating MSSM but is unique to the BMSSM with CP violation. In the BMSSM without CP violation only the two scalar Higgs bosons couple to the weak gauge bosons and in the MSSM with CP violation one Higgs is always light ($M_{H_1}\lesssim 130$~GeV) and therefore has only a small to moderate branching ratio into WW.
Tab.~\ref{tab:scenarios} shows the predictions for the Higgs spectrum, effective Higgs - ZZ and Higgs - gg couplings as well as the most important Higgs branching ratios for the example parameter set given in the first row of  Tab.~\ref{tab:scenarios_input} that realizes such a scenario. Fig.~\ref{fig:sce1} shows these quantities as functions of the phase of $\alpha$, keeping the phase of $\omega$ fixed to the same ratio Arg$(\omega)$/Arg$(\alpha) = -1/5$ as in  the first row of Tab.~\ref{tab:scenarios_input}.

In general, both in this as well as in the other scenarios considered below, one observes a strong dependence of the Higgs couplings and branching ratios on the phase of $\alpha$, accompanied by a non-negligible dependence of the mass spectrum as well. For non trivial values of Arg$(\alpha)$ all three Higgs boson couple to the gauge bosons and all three couplings can be large enough that all three Higgs bosons decay dominantly into $WW$ (see for example Arg$(\alpha)\sim 1.5$).

Concerning the spectrum, we remark that for $-1 \lesssim$ Arg$(\alpha) \lesssim 1$, the lightest Higgs boson lies between $160$~GeV and $165$~GeV, a region where the SM Higgs is excluded by the Tevatron. Due to the fact that in our scenario this Higgs decays mainly into $WW$ and has a production cross section that is only slightly reduced compared to the SM case, Tevatron data indeed excludes small values of Arg$(\alpha)$ in this scenario.
In addition, we observe that the heaviest Higgs is around $200$~GeV in the full range for Arg$(\alpha)$ and therefore has a sizable branching fraction not only to $WW$ but also to $ZZ$. Correspondingly, such a heaviest Higgs would first be detected in the $H_3 \to ZZ \to 4\ell$ channel at the LHC.

We remark however, that scenarios in which all three Higgs bosons decay dominantly to $WW$ do not necessarily imply a heavy Higgs with a mass of $M_{H_3} \gtrsim 200$~GeV. We also found scenarios with all Higgs masses in the range $155$~GeV - $180$~GeV, where the main decay channel for all Higgs bosons is $H_i \to WW$.

\paragraph{Scenario Ib:}

\begin{figure}[H] \centering
\includegraphics[width=0.39\textwidth]{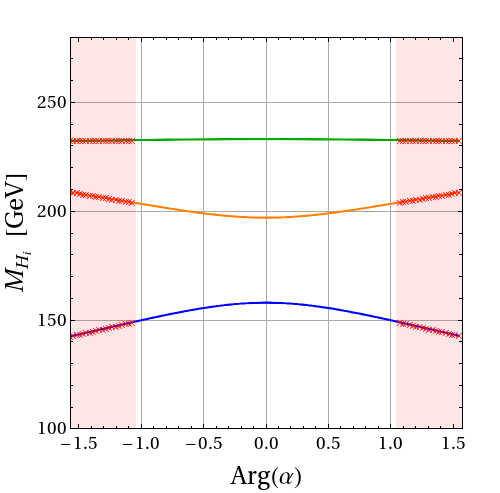} ~~~
\includegraphics[width=0.39\textwidth]{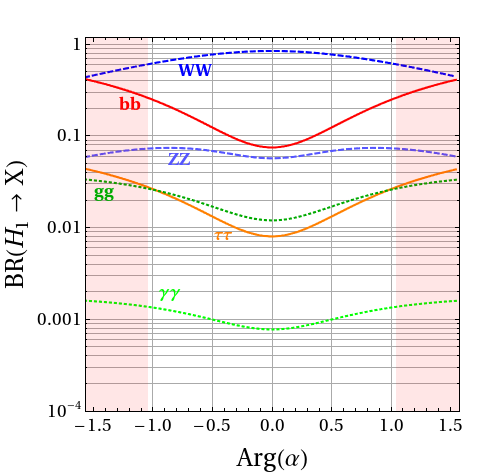} \\
\includegraphics[width=0.39\textwidth]{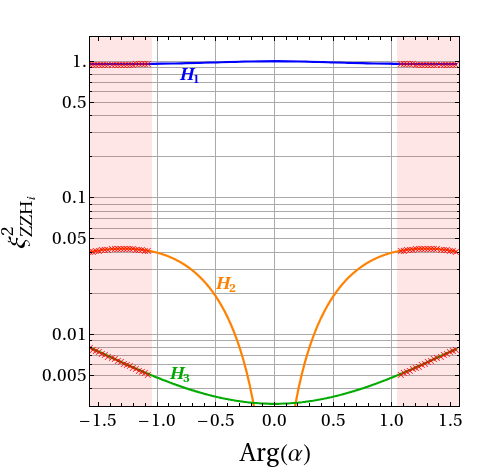} ~~~
\includegraphics[width=0.39\textwidth]{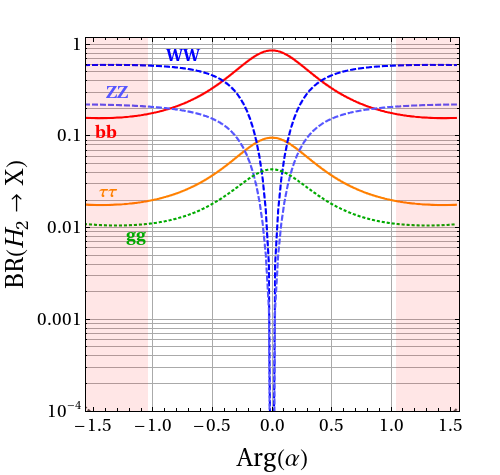} \\
\includegraphics[width=0.39\textwidth]{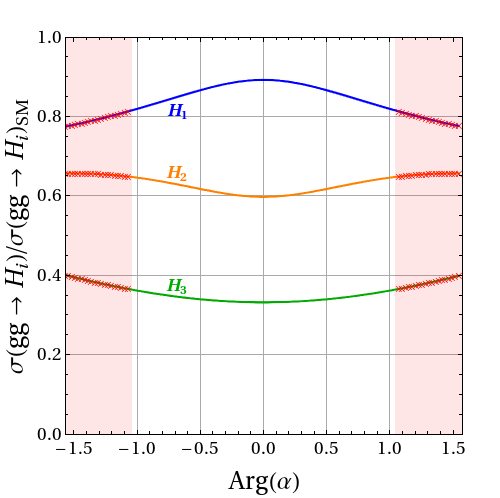} ~~~
\includegraphics[width=0.39\textwidth]{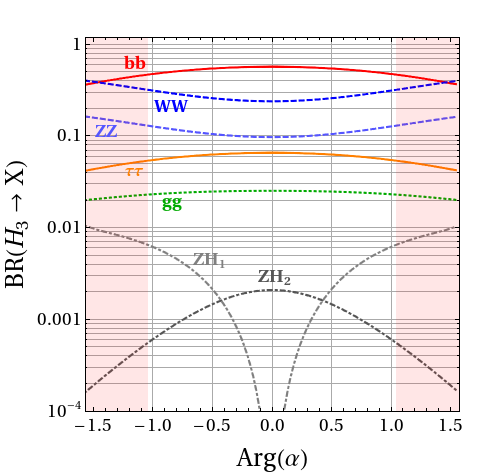}
\caption{\small 
Higgs spectrum, Higgs - vector boson couplings, $gg \to H_i$ production cross sections and Higgs branching ratios as function of the phase of $\alpha$ for the Scenario Ib. The phase of $\omega$ is fixed to Arg$(\omega) = -$Arg$(\alpha)/5$ in the plots, to keep under control EDM constraints. The side bands (light red shaded regions) correspond to vacua that are only local minima of the potential.}
\label{fig:sce3}
\end{figure}

This scenario consists of a heavy SM like Higgs with a mass of $M_{H_1} \simeq 150$~GeV that decays mainly into $WW$, and two additional heavy Higgs bosons with masses $M_{H_{2,3}} \gtrsim 200$~GeV. 
An example input parameter set for such a scenario can be found in the second row of Tab.~\ref{tab:scenarios_input}. In Tab.~\ref{tab:scenarios} the corresponding Higgs spectrum, effective Higgs - ZZ and Higgs - gg couplings as well as the most important Higgs branching ratios are shown. Finally, Fig.~\ref{fig:sce3} shows them as functions of the phase of $\alpha$, for the same ratio Arg$(\omega)$/Arg$(\alpha) = -1/5$ as in the second row in Tab.~\ref{tab:scenarios_input}.

From the plots in Fig.~\ref{fig:sce3} we observe that the $H_1 \to WW$ decay is indeed the main channel for the lightest Higgs for the full allowed region of Arg$(\alpha)$. The largest branching fraction for the heaviest Higgs on the other hand is always $H_3 \to b\bar b$. While the heaviest Higgs mainly corresponds to the heavy scalar of the theory, its coupling to gauge bosons is tiny. This happens because this scenario already approaches the decoupling limit where the lightest Higgs is SM like and carries basically the full coupling to gauge bosons. In absence of CP violation, the second heaviest Higgs is the pseudoscalar and therefore does not couple at all to gauge bosons. For non-trivial values of the phase of $\alpha$ however (see e.g. Arg$(\alpha) \simeq -1$), the would-be pseudoscalar can mix sufficiently strong with the light scalar, acquiring a non-negligible coupling to gauge bosons and decaying mainly into $WW$.

As the lightest Higgs has a mass of about $(150-155)$~GeV and is SM like, it might be even in the reach of Tevatron. Also the second Higgs decays mainly into gauge bosons and the best search channel at LHC is $H_2 \to ZZ \to 4\ell$.
Even though the heaviest Higgs has the largest branching ratio into $b\bar b$ we expect the most promising search channel to be $H_3 \to ZZ \to 4\ell$, due to the non-negligible branching ratio into $ZZ$ at the level of 10\%.
Still very high statistics will be required to probe this Higgs boson.

\paragraph{Scenario II:}

\begin{figure}[H] \centering
\includegraphics[width=0.39\textwidth]{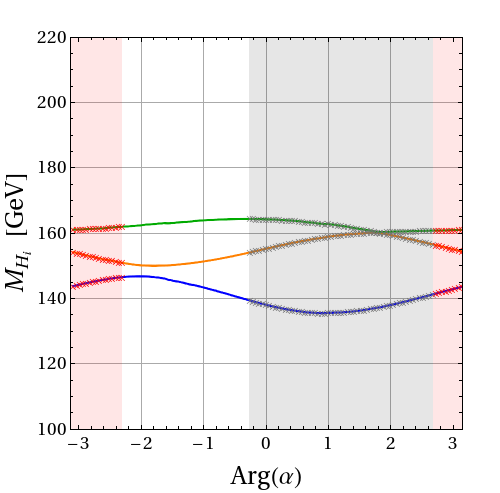} ~~~
\includegraphics[width=0.39\textwidth]{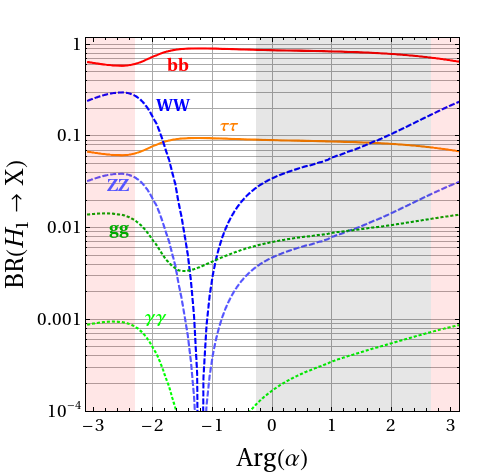} \\
\includegraphics[width=0.39\textwidth]{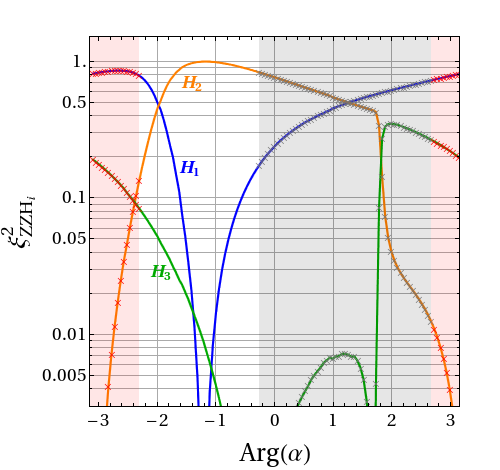} ~~~
\includegraphics[width=0.39\textwidth]{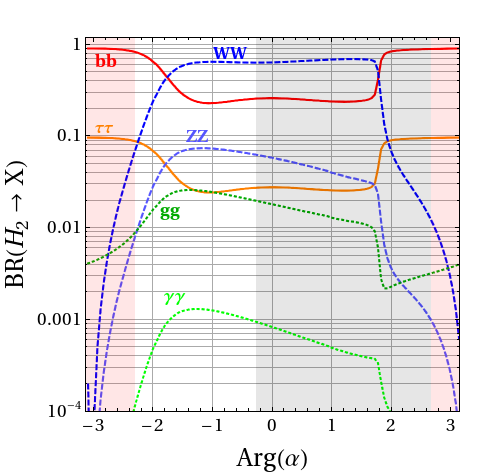} \\
\includegraphics[width=0.39\textwidth]{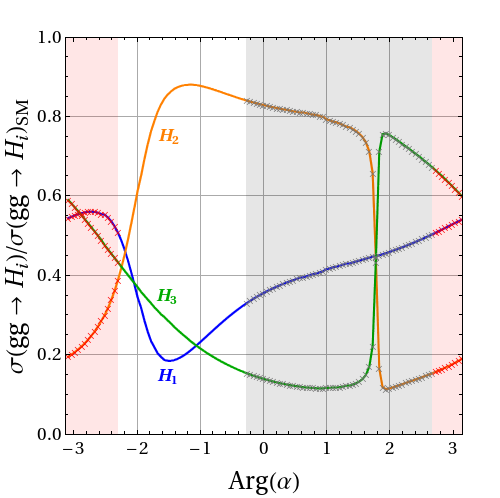} ~~~
\includegraphics[width=0.39\textwidth]{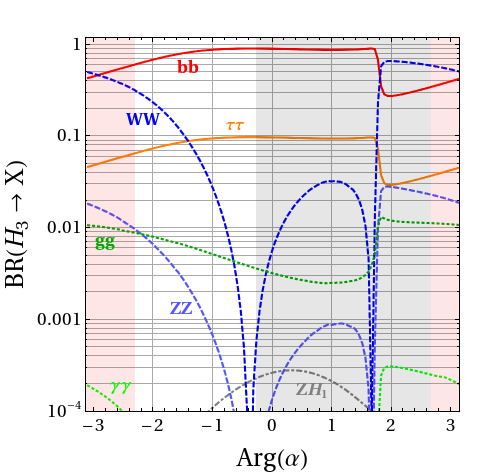}
\caption{\small 
Higgs spectrum, Higgs - vector boson couplings, $gg \to H_i$ production cross sections and Higgs branching ratios as function of the phase of $\alpha$ for the Scenario II. The phase of $\omega$ is fixed to Arg$(\omega) = \pi/20$ in the plots, to keep under control EDM constraints. The inner band (light gray shaded region) is excluded by EDM constraints. The external bands (light red shaded regions) correspond to vacua that are only local minima of the potential.}
\label{fig:sce2}
\end{figure}

In this scenario all three Higgs bosons have similar masses in the range $145$~GeV $\lesssim M_{H_i} \lesssim 160$~GeV and one would naively expect them to decay dominantly into $WW$. However the situation is reversed compared to Scenario Ia discussed above, since all the three Higgs bosons are decaying mainly into $b \bar b$.
An example parameter point that leads to such a scenario is given in the third row of Tab.~\ref{tab:scenarios_input}. Its Higgs spectrum, effective Higgs - ZZ and Higgs - gg couplings as well as the most important Higgs branching ratios are again summarized in Tab.~\ref{tab:scenarios} and also shown as function of the phase of $\alpha$ in Fig.~\ref{fig:sce2}, with Arg$(\omega)$ fixed to $\pi/20$.

From the plots in Fig.~\ref{fig:sce2} we observe that the $H_i \to b\bar b$ decay is indeed the main channel for the three Higgs bosons in the region close to Arg$(\alpha)\sim -2$.
While for non trivial phases the $H_iWW$ coupling is again shared among all three Higgs bosons, this time it is made {\it small} enough such that the dominant decay mode for all three Higgs bosons is $H_i \to b\bar b$.
Correspondingly such a scenario is essentially unconstrained by Tevatron. 

Concerning the Higgs searches at the 7~TeV LHC, probing this scenario appears to be very challenging. 
As it is evident from Tab.~\ref{tab:scenarios}, the $gg \to H_i \to \gamma\gamma$ cross sections are strongly suppressed compared to the SM by a factor of 5 - 15. Even though the $V \to VH_i \to V\tau\tau$ channels are enhanced compared to the SM by a factor of 1.5 - 2, the large masses of the Higgs bosons in the considered scenario prevent the Higgs bosons to be probed in this search mode. A similar conclusion is expected for the boosted Higgs search. Therefore, despite the suppressed branching ratios, the most promising channel seems to be still $H_i \to WW$, even if large statistics will be required in order to be sensitive to the small signal that the scenario predicts.

\pagebreak
\begin{table}[H] 
\addtolength{\arraycolsep}{10pt}
\renewcommand{\arraystretch}{1.3}
\centering
\begin{tabular}{|l|c|c|c|}
\hline\hline
Scenario Ia & $H_1$ & $H_2$ & $H_3$ \\ 
\hline\hline
$M_{H_i}$ [GeV] & 157 & 177 & 202  \\ \hline
$\xi^2_{ZZH_i}$ & 0.94 & 0.04 & 0.02  \\ \hline
$\xi^2_{ggH_i}$ & 0.72 & 0.62 & 0.47  \\ \hline
BR$(H_i \to bb)$ & 15\% (8\%) & 34\% (0.6\%) & 24\% (0.2\%) \\
BR$(H_i \to WW)$ & 76\% (83\%) & 58\% (95\%) & 53\% (74\%) \\
BR$(H_i \to ZZ)$ & 6\% (7\%) & 2\% (4\%) & 19\% (26\%) \\
BR$(H_i \to \gamma\gamma) \times 10^4$ & 9 (9) & 0.8 (1.2) & 0.2 (0.5) \\
\hline\hline
\end{tabular}
\\[10pt]
\begin{tabular}{|l|c|c|c|}
\hline\hline
Scenario Ib & $H_1$ & $H_2$ & $H_3$ \\ 
\hline\hline
$M_{H_i}$ [GeV] & 153 & 201 & 233  \\ \hline
$\xi^2_{ZZH_i}$ & 0.96 & 0.03 & 0.004  \\ \hline
$\xi^2_{ggH_i}$ & 0.84 & 0.64 & 0.35  \\ \hline
BR$(H_i \to bb)$ & 19\% (13\%) & 21\% (0.2\%) & 51\% (0.1\%) \\
BR$(H_i \to WW)$ & 69\% (74\%) & 56\% (74\%) & 29\% (71\%) \\
BR$(H_i \to ZZ)$ & 7\% (8\%) & 19\% (26\%) & 12\% (29\%) \\
BR$(H_i \to \gamma\gamma) \times 10^4$ & 12 (12) & 0.6 (0.5) & 0.5 (0.3) \\
\hline\hline
\end{tabular}
\\[10pt]
\begin{tabular}{|l|c|c|c|}
\hline\hline
Scenario II & $H_1$ & $H_2$ & $H_3$ \\ 
\hline\hline
$M_{H_i}$ [GeV] & 147 & 150 & 162  \\ \hline
$\xi^2_{ZZH_i}$ & 0.62 & 0.32 & 0.06  \\ \hline
$\xi^2_{ggH_i}$ & 0.41 & 0.53 & 0.39  \\ \hline
BR$(H_i \to bb)$ & 69\% (22\%) & 72\% (16\%) & 65\% (2\%)  \\
BR$(H_i \to WW)$ & 20\% (63\%) & 17\% (69\%) & 26\% (94\%)  \\
BR$(H_i \to ZZ)$ & 3\% (8\%) & 2\% (8\%) & 1\% (3\%)  \\
BR$(H_i \to \gamma\gamma) \times 10^4$ & 6 (16) & 3 (13) & 0.5 (4) \\
\hline\hline
\end{tabular}
\\[10pt]
\caption{\small
Predictions for the Higgs spectrum, the effective $ZZH_i$ and $ggH_i$ couplings as well as the branching ratios for the most important decay channels of the Higgs bosons in the example scenarios defined in Tab.~\ref{tab:scenarios_input}. 
The corresponding SM branching ratios for a SM Higgs with the same mass are given in parenthesis. In all scenarios of this section the $H_i \to \tau \tau$ branching ratios are given approximately by BR$(H_i \to \tau \tau) \simeq  \frac{1}{10}$BR$(H_i \to bb)$.
}
\label{tab:scenarios}
\end{table}

\begin{figure}[t] \centering
\includegraphics[width=0.45\textwidth]{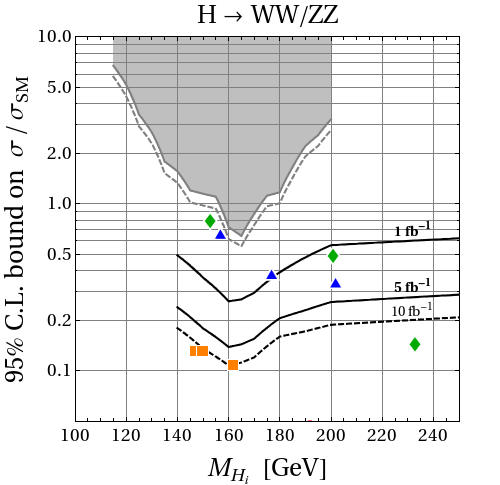} ~~~
\includegraphics[width=0.45\textwidth]{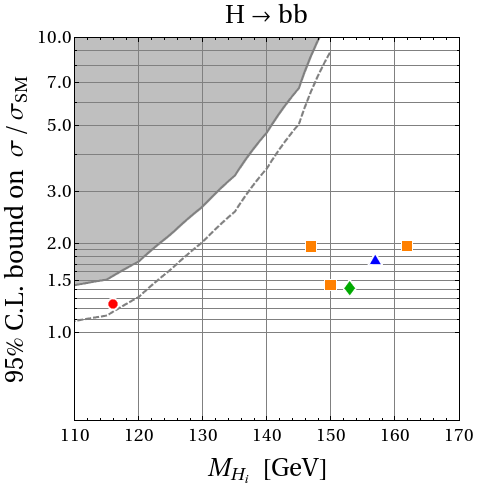}
\caption{\small 
Left: 95\% C.L. sensitivities of Tevatron and LHC in the $H \to WW / ZZ$ channels.  Right: 95\% C.L. sensitivities of Tevatron in the $H \to b \bar b$ decays channels with the Higgs produced in association with a vector boson. The black lines in the left plot are the combined expected ATLAS and CMS sensitivities for 1, 5, and 10~fb$^{-1}$. The gray region in both plots is excluded by Tevatron, the dashed gray line shows a naive extrapolation of the Tevatron sensitivity with 10~fb$^{-1}$.
The blue triangles, green diamonds and orange squares correspond to the Higgs bosons of Scenario Ia, Ib and II, respectively. The red circle in the right plot corresponds to the lightest Higgs boson in a scenario with large $B_s$ mixing phase to be discussed in Sec.~\ref{sec:Bsmixing}.
}
\label{fig:collider_sensitivity}
\end{figure}

Fig.~\ref{fig:collider_sensitivity} shows the Tevatron and LHC sensitivities to the three scenarios discussed above. The blue triangles, green diamonds and orange squares correspond to the Higgs bosons of Scenario Ia, Ib and II, respectively.
The left plot shows the product of the $gg \to H_i$ production cross section times the branching ratio into vector bosons normalized to the SM value. The gray region is excluded at the 95\% confidence level by current Tevatron data and is based on a naive combination of CDF and D0 bounds~\cite{CDFhiggs6,Abazov:2011bc,D0higgs3}. We mention that the $H_i \to WW/ZZ$ results in~\cite{CDFhiggs6,Abazov:2011bc,D0higgs3} are not only based on gluon-gluon fusion but contain also vector boson fusion and associated production at a level of up to 25\%. Still, as the $gg \to H_i \to WW/ZZ$ channel dominates the obtained constraints, we consider the shown bound to be a reasonable estimate of the combined Tevatron exclusion reach. The gray dashed line shows the expected Tevatron sensitivity, extrapolating the CDF and D0 bounds to an integrated luminosity of 10~fb$^{-1}$ each. To be conservative, for the extrapolation we only take into account the higher statistics and do not assume any additional improvements of the analysis.
The black lines correspond to a combination of the expected sensitivities from ATLAS~\cite{ATLAS} and CMS~\cite{CMS} at a center of mass energy of 7 TeV. The ATLAS and CMS curves are based on a combination of several production mechanisms. However, in the shown region with $M_{H_i} > 140$~GeV, the dominant production mechanism is $gg \to H_i$, with vector boson fusion and associated production contributing only $\sim 10$\% and $\sim 5$\%, respectively (see e.g. Tab.~2 in~\cite{ATLAS}).

We observe that the lightest Higgs bosons of Scenarios Ia and Ib are close to the current Tevatron exclusion and just outside the final expected Tevatron sensitivity. They should be easily accessible at the LHC already with an integrated luminosity of 1~fb$^{-1}$.
As expected, all three Higgs bosons of Scenario Ia are well within the LHC sensitivity and can be probed with 5~fb$^{-1}$. In Scenario Ib on the other hand, only the two lightest Higgs bosons can be probed in the $H_i \to WW$ channel, while the heaviest Higgs that decays mainly into $b \bar b$ is inaccessible in the 7~TeV LHC run, unless more than 20~fb$^{-1}$ per experiment are collected.

In Scenario II finally, despite their large masses, all three Higgs bosons decay mainly into $b \bar b$ and are correspondingly harder to be probed in the $H_i \to WW$ channel. Combining the sensitivities of both ATLAS and CMS, one would need more than 10~fb$^{-1}$ at the 7~TeV LHC to probe the three Higgs bosons in this scenario.\footnote{
The two lightest Higgs bosons of Scenario II are very close in mass with a mass splitting of only 3~GeV. Given the finite mass resolution of ATLAS and CMS, these two Higgs bosons could appear as a single particle with the combined cross section and thus might be accessible in the particular scenario considered here.}

The right plot of Fig.~\ref{fig:collider_sensitivity} shows the Tevatron sensitivity to the Higgs bosons of the three scenarios in the production in association with a vector boson and the subsequent decay into a $b \bar b$ pair. The gray region is excluded by current Tevatron data at the 95\% confidence level and based on a naive combination of CDF and D0 results~\cite{CDFhiggs1,CDFhiggs2,CDFhiggs3,CDFhiggs4,CDFhiggs5,Abazov:2010hn,D0higgs1,D0higgs2}. We checked that our combination roughly approximates the official Tevatron combination in the low mass region~\cite{:2010ar}.
The dashed gray line shows a conservative extrapolation of the Tevatron sensitivity with 10~fb$^{-1}$ from both CDF and D0, taking into account only the higher statistics in the extrapolation.
Even though several Higgs bosons of our scenarios show a signal that is enhanced by a factor of 1.5 - 2 compared to the SM, we observe that, due to their rather heavy masses, they are far outside the expected Tevatron bounds.
The same is true for the $WW \to H_i \to \tau^+ \tau^-$ mode and the boosted Higgs searches at the 7~TeV LHC.

\section{The \boldmath{\texorpdfstring{$B_s$}{Bs}} Mixing Phase in the BMSSM} \label{sec:Bsmixing}

In this section we study the possible impact of the new CP phases in the Higgs sector of the BMSSM on CP violation in meson mixing. 

The $B_s$ mixing phase is predicted to be tiny in the Standard Model and therefore offers excellent opportunity to probe new sources of CP violation in NP models. Interestingly, combining data from CDF and D0 on the time dependent CP asymmetry in $B_s \to \psi\phi$~\cite{Aaltonen:2007he,:2008fj,CDFSpsiphi,D0Spsiphi} as well as a D0 analysis of the like-sign dimuon charge asymmetry~\cite{Abazov:2010hv}, one finds a large $B_s$ mixing phase at the level of (2-3)$\sigma$ above the SM prediction~\cite{Ligeti:2010ia,Lenz:2010gu}. A very recent update of the D0 analysis~\cite{Abazov:2011yk} finds a discrepancy in the like-sign dimuon charge asymmetry of 3.9$\sigma$ with respect to the SM and strengthens this result.
In addition, tensions in the fits of the CKM matrix seem to hint towards sizable NP contributions 
to CP violation in $B_d$ mixing~\cite{Lunghi:2008aa,Buras:2008nn,Lenz:2010gu,Lunghi:2010gv}.

As shown in~\cite{Buras:2010mh,Buras:2010zm}, generic two Higgs doublet models with Minimal Flavor Violation (MFV), where the CKM matrix is the only source of flavor violation~\cite{Chivukula:1987py,Buras:2000dm,D'Ambrosio:2002ex} and additional CP violating phases in the Yukawa and the Higgs sector are allowed, are able to generate large effects both in the $B_s$ and $B_d$ mixing phases in agreement with the present experimental data and compatible with constraints from BR$(B_{s,d}\to\mu^+\mu^-)$ and EDMs.
The same is not true in the MSSM with MFV. While 1-loop box contributions to meson mixing are generically small~\cite{Altmannshofer:2007cs}, sizable effects can in principle be generated at the 2-loop level in the large $\tan\beta$ regime by so called double Higgs penguin contributions~\cite{Isidori:2001fv,Buras:2002vd}. However, these double Higgs penguins are strongly constrained by the experimental bound on the branching ratio of the rare $B_s \to \mu^+ \mu^-$ decay~\cite{Carena:2006ai}. As a result one finds that CP violation in meson mixing remains SM like in the MSSM with MFV~\cite{Altmannshofer:2008hc,Altmannshofer:2009ne}\footnote{A sizable $B_s$ mixing phase can be possible in the so-called uplifted SUSY Higgs region~\cite{Dobrescu:2010mk,Dobrescu:2010rh}, even though such a framework is strongly constrained by $B$ physics observables and $(g-2)_\mu$~\cite{Altmannshofer:2010zt}.} and additional sources of flavor violation are required to generate large NP effects in meson mixing phases.\footnote{
SUSY models that contain sources of flavor violation beyond the CKM matrix and that are capable of generating a sizable $B_s$ mixing phase have been studied for example in~\cite{Dutta:2008xg,Ko:2008xb,Giudice:2008uk,Altmannshofer:2009ne,Parry:2010ce,King:2010np,Buras:2010pm,Girrbach:2011an,Barbieri:2011ci,Crivellin:2011sj}.}

In the following we analyze to which extent this result is changed in the BMSSM. We consider a minimal flavor violating soft sector, i.e. no new sources of flavor violation in addition to the CKM matrix and investigate the impact of the higher dimensional operators with new sources of CP violation on B physics observables.

\subsection{Basics of Meson Mixing}

The $B_q$ mixing amplitude is given by 
\begin{equation}
\langle B_q | \mathcal{H}_\textnormal{eff}| \bar B_q \rangle = M^q_{12} - \frac{i}{2}\Gamma^q_{12}~.
\end{equation}
Assuming no NP effects in the absorptive part $\Gamma_{12}^q$ that is dominated by SM tree level contributions, the effects of new short distance dynamics can be conveniently parametrized by
\begin{equation} \label{eq:oioi}
M_{12}^q = (M_{12}^q)_{\rm SM} + (M_{12}^q)_{\rm NP} = C_{B_q} e^{i 2\theta_q} (M_{12}^q)_{\rm SM}~.
\end{equation}
The $B_{d,s}$ mass differences and the CP asymmetries $S_{\psi K_S}$ and $S_{\psi\phi}$ in the $B_d \to \psi K_S$ and $B_s \to \psi\phi$ decays are then given by
\begin{eqnarray}
\Delta M_q &=& 2\left|M_{12}^q\right| = (\Delta M_q)_{\text{SM}}C_{B_q}~, \label{eq:Delta_Mq} \\
S_{\psi K_S} &=& \sin( \textnormal{Arg}(M_{12}^d)) = \sin( 2\beta + 2\theta_d )~,\\
S_{\psi\phi} &=& - \sin( \textnormal{Arg}(M_{12}^s)) = \sin( 2|\beta_s| - 2\theta_s )~, \label{eq:CPV_Bq}
\end{eqnarray}
where the SM angles $\beta$ and $\beta_s$ are the phases of the CKM elements $V_{td}$ and $V_{ts}$ in the standard CKM phase convention
\begin{equation}
V_{td} = |V_{td}| e^{-i\beta} ~~,~~~ V_{ts} = -|V_{ts}| e^{-i\beta_s} ~.
\end{equation}
While the SM predictions for the mass differences are in excellent agreement with the experimental determinations, the same is not true for the mixing phases. Using as input the CKM parameters from the NP fit in~\cite{Lenz:2010gu} one finds
\begin{equation}
S_{\psi K_S}^\textnormal{SM} = 0.81^{+0.02}_{-0.07} ~~,~~~ S_{\psi\phi}^\textnormal{SM} = 0.044^{+0.002}_{-0.003} ~.
\end{equation}
This should be compared to the measured values~\cite{Asner:2010qj,Lenz:2011ti}
\begin{equation}
S_{\psi K_S}^\textnormal{exp} = 0.67 \pm 0.02 ~~,~~~ S_{\psi\phi} = 0.78^{+0.12}_{-0.19} ~,
\end{equation}
that differ by (2-3)$\sigma$ from the SM predictions.
\footnote{Note that the result for $S_{\psi\phi}$ from~\cite{Lenz:2011ti} does not include yet the updates on the time dependent CP asymmetries in $B_s \to \psi\phi$ from CDF and D0~\cite{CDFSpsiphi,D0Spsiphi}. As these updates are in better agreement with the SM prediction, including them is expected to slightly decrease the central value for $S_{\psi\phi}$.}

\subsection[BMSSM Contributions to \texorpdfstring{$B$}{B} Meson Mixing]{\boldmath BMSSM Contributions to $B$ Meson Mixing}

New Physics contributions to the $B_q$ mixing amplitude can be encoded in the effective Hamiltonian
\begin{equation}
\mathcal{H}_\textnormal{eff}^{\rm NP} = - \sum_{i=1}^{5} C_i O_i - \sum_{i=1}^{3} \tilde C_i \tilde O_i ~,
\end{equation}
with the operators $O_i$ and $\tilde O_i$ given e.g. in~\cite{Ciuchini:1997bw,Buras:2000if}.

In the BMSSM with no new sources of flavor violation in addition to the CKM matrix, sizable contributions to the mixing amplitudes can be generated in the large $\tan\beta$ regime. The two most important operators that contribute to $B_q$ mixing in such a framework read
\begin{equation}\label{eq:operatorsDF2}
\tilde O_2 = (\bar b_R q_L)(\bar b_R q_L) ~,~~ O_4 = (\bar b_R q_L)(\bar b_L q_R) ~.
\end{equation}
To evaluate the corresponding contributions to the mixing amplitudes, renormalization group evolution~\cite{Ciuchini:1997bw,Buras:2000if} is used to run down the corresponding Wilson coefficients $C_4$ and $\tilde C_2$ from the matching scale to the scale of the $B$ mesons where the hadronic operator matrix elements are given~\cite{Becirevic:2001xt}.

In the following we discuss the contributions to the Wilson coefficients, showing the leading terms in $\tan\beta$ and performing a $v/\tilde m$ expansion. However, in our numerical analysis we implement these contributions following the procedure described in~\cite{Buras:2002vd} that goes beyond these approximations.

\begin{figure}[t] \centering
\includegraphics[width=0.9\textwidth]{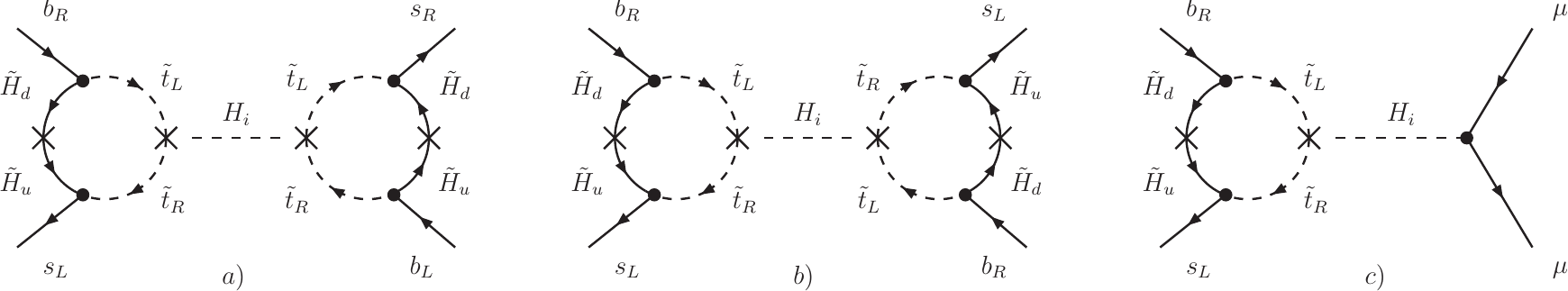}
\caption{
Diagrams a) and b) show the most important double Higgs penguin contributions to $B_s$ mixing in our framework. Diagram c) is the dominant contribution to $B_s \to \mu^+\mu^-$. The leading $\tan\beta$ enhanced contributions to the mixing and decay amplitudes come from flavor changing self-energy corrections in diagrams where the Higgs propagators are attached to the external quark legs.
}
\label{fig:BPhisics_diagrams}
\end{figure}

The Wilson coefficients of the above operators are mainly generated by double Higgs penguin contributions. In the case of $B_s$ mixing, the dominant Higgsino loop diagrams are shown in Fig.~\ref{fig:BPhisics_diagrams} and can be approximated by
\begin{eqnarray} \label{eq:C4_DP_H}
C_4^{\tilde H} &=& \frac{\alpha_2^3}{4\pi} \frac{m_b m_s}{M_W^2} \frac{t_\beta^4}{|1+\epsilon_b t_\beta|^2|1+\epsilon_0 t_\beta|^2} \frac{|\mu A_t|^2}{\tilde m^4} \frac{m_t^4}{16 M_W^4} \big( f_1(x_\mu)\big)^2 ~ (V_{tb} V_{ts}^*)^2  \nonumber \\
&& ~~~ \times \sum_{i=1}^3 \frac{1}{M_{H_i}^2} \left[ \left( \frac{c_\alpha}{s_\beta} O_{2i} - \frac{s_\alpha}{s_\beta} O_{1i} \right)^2 + O_{3i}^2 \right] ~, \\  \label{eq:C2_DP_H}
\tilde C_2^{\tilde H} &=& \frac{\alpha_2^3}{4\pi} \frac{m_b^2}{M_W^2} \frac{t_\beta^4}{(1+\epsilon_b^* t_\beta)^2(1+\epsilon_0^* t_\beta)^2} \frac{(\mu A_t)^2 e^{2 i \theta}}{\tilde m^4} \frac{m_t^4}{32 M_W^4} \big( f_1(x_\mu)\big)^2 ~ (V_{tb} V_{ts}^*)^2  \nonumber \\
&& ~~~ \times \sum_{i=1}^3 \frac{1}{M_{H_i}^2} \left[ \left( \frac{c_\alpha}{s_\beta} O_{2i} - \frac{s_\alpha}{s_\beta} O_{1i} \right)^2 - O_{3i}^2 - 2i O_{3i} \left( \frac{c_\alpha}{s_\beta} O_{2i} - \frac{s_\alpha}{s_\beta} O_{1i} \right)\right] ~. 
\end{eqnarray}
These expressions assume a common mass $\tilde m$ for the squarks. MFV frameworks in principle allow for a mass splitting between the first two and the third generation of left-handed squarks that can be induced radiatively through the Yukawa couplings. In such a case, in addition to the Higgsino contributions shown here, also gluino contributions to FCNC processes arise and can become relevant~\cite{Bertolini:1990if,Carena:2008ue}. They are consistently included in our numerical analysis.
The $\tan\beta$ resummation factors appearing in~(\ref{eq:C4_DP_H}) and~(\ref{eq:C2_DP_H}) can be approximated by
\begin{equation}
\epsilon_b \simeq \epsilon^{\tilde g} + \epsilon^{\tilde H}+ \epsilon^{\tilde W}~,~~~ \epsilon_0 \simeq \epsilon^{\tilde g} + \epsilon^{\tilde W}~,
\end{equation}
with the gluino, Higgsino and Wino contributions given already in Eq.~(\ref{eq:epsilon}). The loop function $f_1$ is given in appendix~\ref{app:loop}.

Expanding the Higgs propagators in (\ref{eq:C4_DP_H}) and (\ref{eq:C2_DP_H}) in $M_Z^2/M_A^2$, $\cot\beta$ and $1/M$ we find in agreement with~\cite{Gorbahn:2009pp}
\begin{eqnarray} 
\sum_{i=1}^3 \frac{1}{M_{H_i}^2} \left[ \left( \frac{c_\alpha}{s_\beta} O_{2i} - \frac{s_\alpha}{s_\beta} O_{1i} \right)^2 + O_{3i}^2 \right] &\simeq& \frac{2}{M_A^2} ~\Big( 1 + O(1/M) \Big) ~,
\end{eqnarray}
\begin{eqnarray}
\sum_{i=1}^3 \frac{1}{M_{H_i}^2} \left[ \left( \frac{c_\alpha}{s_\beta} O_{2i} - \frac{s_\alpha}{s_\beta} O_{1i} \right)^2 - O_{3i}^2 - 2i O_{3i} \left( \frac{c_\alpha}{s_\beta} O_{2i} - \frac{s_\alpha}{s_\beta} O_{1i} \right)\right] &\simeq& \nonumber \\ \simeq ~ - \frac{v^2 \lambda_5^* e^{- 2 i \theta}}{M_A^4} + \frac{v^4 (\lambda_6^*)^2 e^{- 2 i \theta}}{M_A^4 M_h^2} && ~.
\end{eqnarray}
Some comments are in order: the contribution to $C_4$ is proportional to $m_s m_b$ for $B_s$ mixing and to $m_d m_b$ in case of $B_d$ mixing and therefore completely negligible in the latter case. The phase of the contribution to $C_4$ is given by the same tiny phase appearing in the SM contribution $2$Arg$(V_{ts}^*) = - 2\beta_s \simeq 2^\circ$, even for complex $\mu$ and $A_t$. This is true in the approximation that enters the above equation that the squarks are all degenerate with a mass $\tilde m$.\footnote{For large splittings between the strange and the bottom squark masses also $C_4$ can in principle depend on NP phases~\cite{Carena:2006ai,Hofer:2009xb,Dobrescu:2010rh}.} Compared to the MSSM, the contributions to $C_4$ receive only {\it corrections} at the $1/M$ level.

The contribution to $\tilde C_2$ is proportional to $m_b^2$ both for $B_s$ and $B_d$ mixing, leading to NP effects in $B_s$ and $B_d$ mixing that are comparable in size. In the absence of the higher dimensional operators in the Higgs sector, i.e. in the MSSM, the contribution to $\tilde C_2$ is highly suppressed by $\cot^2\beta M_W^2/M_A^2$ and, despite its enhancement by $m_b/m_q$ compared to $C_4$, it is usually completely irrelevant~\cite{Buras:2002vd,Carena:2006ai,Gorbahn:2009pp}.
In the BMSSM on the other hand, once $1/M$ effects are taken into account, $\tilde C_2$ has the same $\tan^4\beta$ dependence as $C_4$ and is only suppressed by $1/M$ and $v^2/M_A^2$. Correspondingly, due to its enhancement by $m_b/m_q$, it is typically much more important than $C_4$, in particular for small Higgs masses.
In addition, the contribution to $\tilde C_2$ is very sensitive to NP sources of CP violation. Its phase, that in turn generates a phase of the mixing amplitude $M_{12}$ through~(\ref{eq:oioi}), can be induced by $\mu A_t$, the phase of the Higgs VEV $\theta$, or directly through the Higgs propagator by the parameters $\lambda_5$ and $\lambda_6$.
A sizable complex $\tilde C_2$ is the main {\it qualitative difference} between the BMSSM and the MSSM in the flavor sector.

\subsection{Constraints from other Flavor Observables}

There are several severe flavor constraints on MFV frameworks in the large $\tan\beta$ regime \cite{Isidori:2006pk,Altmannshofer:2010zt}. In our case, the most important ones are the rare $B_s \to \mu^+ \mu^-$ and $B \to X_s \gamma$ decays. While also the $B \to \tau \nu$, $B \to D \tau \nu$ and $K \to \mu \nu$ decays are known to be very sensitive probes of extended Higgs sectors, they are much less constraining than $B_s \to \mu^+ \mu^-$ and $B \to X_s \gamma$ in regions of parameter space where the $B_s$ mixing amplitude receives sizable NP contributions.\footnote{
While there is a tension at the $(2-3)\sigma$ level between the experimental determination of BR$(B \to \tau \nu)$ and the low value of $V_{ub}$ that is preferred by fits of the Unitarity Triangle~\cite{Bona:2009cj}, a more conservative SM prediction of the BR$(B \to \tau \nu)$, that is based on the PDG value of $V_{ub}$, still leaves sizable room for NP contributions to $B \to \tau \nu$.
}

\subsubsection[\texorpdfstring{$B_s \to \mu^+ \mu^-$}{Bs -> mu+ mu-}]{\boldmath $B_s \to \mu^+ \mu^-$}

Recently, the CDF collaboration presented the first two sided bound on the branching ratio of the rare decay $B_s \to \mu^+ \mu^-$~\cite{Collaboration:2011fi}. They find
\begin{equation} \label{eq:Bsmm_CDF}
{\rm BR}(B_s \to \mu^+ \mu^-)_{\rm exp} = (1.8^{+1.1}_{-0.9}) \times 10^{-8} ~,
\end{equation}
with a central value that is roughly a factor of six above the SM prediction~\cite{Buras:2003td}
\begin{equation}
{\rm BR}(B_s \to \mu^+ \mu^-)_{\rm SM} = (3.2 \pm 0.2) \times 10^{-9}~.
\end{equation}
CDF also provides an upper bound on the branching ratio at the 95\% C.L.~\cite{Collaboration:2011fi}
\begin{equation} \label{eq:Bsmm_CDF_bound}
{\rm BR}(B_s \to \mu^+ \mu^-)_{\rm exp} < 4.0 \times 10^{-8} ~.
\end{equation}
In our numerical analysis in Sec.~\ref{sec:B_numerics} we will use the upper bound~(\ref{eq:Bsmm_CDF_bound}), and comment on the implications if the central value in~(\ref{eq:Bsmm_CDF}) will be confirmed with larger significance.

As it is well know, the $B_s \to \mu^+ \mu^-$ decay constitutes a very important constraint of the MSSM in the large $\tan\beta$ regime, as its branching ratio grows with $\tan^6\beta$~\cite{Choudhury:1998ze,Babu:1999hn} and enhancements by orders of magnitude compared to the SM prediction are possible. Approximately one has
\begin{equation}
R_{B_s\mu\mu}=\frac{{\rm BR}(B_s \to \mu^+ \mu^-)_{\rm \phantom{SM}}}{{\rm BR}(B_s \to \mu^+ \mu^-)_{\rm SM}}= \left|S\right|^2 + \left|1- P\right|^2 ~.
\end{equation}
The dominant Higgsino contributions to $S$ and $P$ are illustrated in Fig.~\ref{fig:BPhisics_diagrams} and given by
\begin{eqnarray}\label{eq.S}
S^{\tilde H} &\!\!\!=\!\!\!& m_{B_s}^2 \frac{t_\beta^3}{(1+\epsilon_b^* t_\beta)(1+\epsilon_0^* t_\beta)(1+\epsilon_\ell t_\beta)} ~ \frac{m_t^2}{8 M_W^2} \frac{\mu A_t e^{i \theta}}{\tilde m^2} \frac{f_1(x_\mu)}{Y_0(x_t)} \nonumber \\
&&~~~~~~~~~~~~~~~  \times \sum_{i=1}^3 \frac{1}{M_{H_i}^2} \left( \frac{c_\alpha}{s_\beta} O_{2i} - \frac{s_\alpha}{s_\beta} O_{1i} - i O_{3i} \right) \left( \frac{c_\alpha}{s_\beta} O_{2i} - \frac{s_\alpha}{s_\beta} O_{1i} \right) ~, \\\label{eq.P}
P^{\tilde H} &\!\!\!=\!\!\!& m_{B_s}^2 \frac{t_\beta^3}{(1+\epsilon_b^* t_\beta)(1+\epsilon_0^* t_\beta)(1+\epsilon_\ell t_\beta)} ~ \frac{m_t^2}{8 M_W^2} \frac{\mu A_t e^{i \theta}}{\tilde m^2} \frac{f_1(x_\mu)}{Y_0(x_t)} \nonumber \\
&&~~~~~~~~~~~~~~~ \times \sum_{i=1}^3 \frac{1}{M_{H_i}^2} \left( O_{3i} + i \frac{c_\alpha}{s_\beta} O_{2i} - i \frac{s_\alpha}{s_\beta} O_{1i} \right) O_{3i}~.
\end{eqnarray}
For the SM loop function one has $Y_0(x_t) \simeq 0.96$. Gluino contributions to $S$ and $P$ are possible within MFV for a splitting between the first two and the third generation of left-handed squarks~\cite{Carena:2008ue} and are included in our numerical analysis.
In the decoupling limit and at leading order in the $1/M$ expansion, the Higgs propagators entering in the above expressions reduce to
\begin{eqnarray}
\sum_{i=1}^3 \frac{1}{M_{H_i}^2} \left( \frac{c_\alpha}{s_\beta} O_{2i} - \frac{s_\alpha}{s_\beta} O_{1i} - i O_{3i} \right) \left( \frac{c_\alpha}{s_\beta} O_{2i} - \frac{s_\alpha}{s_\beta} O_{1i} \right)  &\simeq& \frac{1}{M_A^2} ~\Big( 1 + O(1/M) \Big) ~,\\
\sum_{i=1}^3 \frac{1}{M_{H_i}^2} \left( O_{3i} + i \frac{c_\alpha}{s_\beta} O_{2i} - i \frac{s_\alpha}{s_\beta} O_{1i} \right) O_{3i} &\simeq& \frac{1}{M_A^2} ~\Big( 1 + O(1/M) \Big)
\end{eqnarray}
and one recovers to first order the well known expressions in the MSSM with MFV. The expressions for $S$ and $P$ in~(\ref{eq.S}) and~(\ref{eq.P}) scale as $\tan^3\beta/M_A^2$ at the amplitude level. As they depend on the same combination of SUSY parameters as the double Higgs penguin contributions to $B_s$ mixing, the NP contributions to $B_s \to \mu^+ \mu^-$ and $B_s$ mixing are strongly correlated in the large $\tan\beta$ regime.

We stress that relative to the MSSM expectations for the branching ratio of the $B_s \to \mu^+ \mu^-$ decay, the BMSSM physics only leads to $1/M$ suppressed corrections, that we however include in our numerical analysis. Analogously, also the $B_d \to \mu^+ \mu^-$ decay remains to first approximation MSSM-like in the BMSSM and, given the current experimental bound on its branching ratio BR$(B_d \to \mu^+ \mu^-) < 6.0 \times 10^{-9}$~\cite{Collaboration:2011fi}, it is much less constraining than the $B_s \to \mu^+ \mu^-$ decay in models with MFV.

\subsubsection[\texorpdfstring{$B \to X_s \gamma$}{B -> Xs gamma}]{\boldmath $B \to X_s \gamma$}

Combining the latest experimental results for the branching ratio of the $B \to X_s \gamma$ decay from~\cite{Asner:2010qj} with the NNLO SM prediction~\cite{Misiak:2006zs} (see also~\cite{Becher:2006pu} and~\cite{Benzke:2010js}) one finds
\begin{equation} \label{eq:Rbsgamma_exp}
R_{bs\gamma} = \frac{{\rm BR}(B \to X_s \gamma)_{\rm \phantom{SM}}}{{\rm BR}(B \to X_s \gamma)_{\rm SM}} = 1.13 \pm 0.12~.
\end{equation}
In frameworks with Minimal Flavor Violation, the prediction for the branching ratio can be approximated by~\cite{Kagan:1998ym,Lunghi:2006hc}
\begin{eqnarray} \label{eq:Rbsgamma}
R_{bs\gamma} &=& 1 + \hat a_{77} \left|C_7^{\rm NP}\right|^2 + \hat a_{88} \left|C_8^{\rm NP}\right|^2 \nonumber \\
&& + {\rm Re}\left( \hat a_7\,C_7^{\rm NP} \right) +{\rm Re}\left( \hat a_8\,C_8^{\rm NP} \right) + {\rm Re}\left( \hat a_{78}  C_7^{\rm NP} C_8^{*\,\rm NP} \right) ~,
\end{eqnarray}
where the coefficients $\hat a_i$ are given by $\hat a_7 = -2.41 + 0.21i$, $\hat a_8 = -0.75 - 0.19i$, $\hat a_{77} = 1.59$, $\hat a_{88} = 0.26$ and $\hat a_{78} = 0.82 - 0.30i$~\cite{Lunghi:2006hc}. The NP contributions to the Wilson coefficients $C_7$ and $C_8$ are evaluated at a scale of $160$~GeV in the above equation.
The Wilson coefficients entering~(\ref{eq:Rbsgamma}) receive the most important contributions from charged Higgs - top loops and Higgsino - stop loops in the scenarios that we consider. Apart from small $\cot\beta/M$ suppressed corrections that are included in our numerical analysis, they have the same form as in the MSSM. At the matching scale they read
\begin{eqnarray} \label{eq:C78_H}
C_{7,8}^{H^\pm} &=& \frac{1-\epsilon_0^* \tan\beta}{1 + \epsilon_b^* \tan\beta} h_{7,8}(y_t)~, \\ \label{eq:C78_c_tanb}
C_{7,8}^{\tilde H} &=& \frac{m_t^2}{\tilde m^2} \frac{\tan\beta}{1 + \epsilon_0^*\tan\beta} \frac{A_t \mu}{\tilde m^2} e^{i \theta} ~f_{7,8}(x_\mu)~,
\end{eqnarray}
with $y_t = m_t^2/M_{H^\pm}^2$ and the loop functions $h_{7,8}$ and $f_{7,8}$ are collected in appendix~\ref{app:loop}.
Gluino contributions to $C_{7,8}$ in the context of MFV are discussed in~\cite{Wick:2008sz,Carena:2008ue} and included in our numerical analysis.
As long as possible phases in the $\tan\beta$ resummation factors do not play an important role, the charged Higgs contributions interfere constructively with the SM contribution. They can be partially canceled by the $\tan\beta$ enhanced chargino contributions if sign$(\mu A_t)~=~+1$. We note the appearance of the phase of the Higgs VEV in the chargino contribution that, as already mentioned in Sec.~\ref{sec:EDMs}, is connected to the $\tan\beta$ enhancement.

\subsection{Numerical Analysis}\label{sec:B_numerics}

Due to the strong correlation between the Higgs penguin contributions to $B_s \to \mu^+ \mu^-$ and the double Higgs penguin contributions to the $B_s$ mixing amplitude, one expects that possible NP effects in $B_s$ mixing are severely constrained by the data on BR$(B_s \to \mu^+ \mu^-)$. We stress however the different behavior of the contributions to the $B_s$ mixing amplitude and $B_s \to \mu^+ \mu^-$ with the Higgs mass and $\tan\beta$. In the decoupling limit one has
\begin{equation}
C_4 \propto \tan^4\beta \frac{1}{M_A^2} ~~,~~~ \tilde C_2 \propto \tan^4\beta \frac{v^2}{M_A^4} ~~~~~\textnormal{and}~~~~~ S,P \propto \tan^3\beta \frac{1}{M_A^2} ~.
\end{equation}
Correspondingly, to keep the BR$(B_s \to \mu^+ \mu^-)$ under control and simultaneously keep a sizable $\tilde C_2$, low values of the Higgs masses and moderate values of $\tan\beta$ appear to be the most promising region of parameter space.

Making $\tan\beta$ as low as possible while keeping a sizable $B_s$ mixing amplitude requires certain choices for the remaining parameters entering the expression of $\tilde C_2$. Apart from the obvious requirements of small Higgs masses, a large $\lambda_5$ and $A_t$, also a sizable negative $\mu$ term increases the size of $\tilde C_2$ significantly through the $\tan\beta$ resummation factors.

The requirement of a large $\mu$ term implies that in the regions of parameter space where a sizable $B_s$ mixing phase is possible, the vacuum is generically not absolutely stable (see discussion in Sec.~\ref{sec:vacuum_stability} and in particular Fig.~\ref{fig:vacuum}). In the minimal BMSSM setup that we considered up to now, typically a very deep second minimum of the Higgs potential arises along the D-flat direction corresponding to a VEV of $v \simeq \sqrt{\mu M / \omega}$. Using simple analytic expressions for the bounce action~\cite{Duncan:1992ai} to estimate the tunneling rate from the electro-weak vacuum to the deeper vacuum, we find that the life times of these electro-weak vacuae are typically much shorter than the age of the universe. This strongly constrains the allowed parameter space and prevents to large extent a large $B_s$ mixing phase. 

However we remark the following: As discussed also in Sec.~\ref{sec:vacuum_stability} this second minimum arises due to negative quartic Higgs couplings that are stabilized by the $1/M^2$ suppressed dimension 6 terms~(\ref{eq:Higgs_potentialnonren}) that in turn are induced by the dimension 5 operator in the superpotential. At the $1/M^2$ level there are however several other operators that can modify the Higgs potential~\cite{Carena:2009gx,Antoniadis:2009rn} and that are not even necessarily suppressed by the same scale $M$. These operators can induce additional dimension 6 terms in the Higgs potential that have strong impact on the vacuum structure for large field values. Adding to the Higgs potential dimension 6 terms as in~(\ref{eq:Higgs_potentialnonren}) with order one coefficients, we checked that they can indeed remove the second minimum as long as their scale is not too large, around $M \simeq 2$~TeV. 
Generically, the additional $1/M^2$ operators also induce corrections to the quartic couplings of the Higgs potential and therefore can affect also the Higgs spectrum in the usual electro-weak vacuum with $v = 246$~GeV. However, these corrections of the quartic couplings depend on combinations of coefficients of the $1/M^2$ operators that are independent of the coefficients entering the induced dimension 6 terms. Correspondingly, in the adopted effective theory approach, one is free to choose the $1/M^2$ corrections to the Higgs potential such that their impact on the quartic couplings is small and simultaneously the electro-weak vacuum is stabilized.

The very light Higgs spectrum together with the large negative $\mu$ term also leads to very large charged Higgs contributions to the $b \to s \gamma$ amplitude. Still, one can be in agreement with the experimental constraint due to the chargino contributions that can (at least partially) cancel the Higgs contribution on condition that sign$(\mu A_t) = +1$. This condition has the advantage that gluino and chargino contributions to the $\tan\beta$ resummation factors add up constructively. Additionally, the mass of the stops should be rather low. We chose to split the masses of the third generation of squarks from the first 2 generations. This opens up the possibility to have also gluino contributions to the FCNC amplitudes which can further enhance $\tilde C_2$ slightly.

\begin{figure}[t] \centering
\includegraphics[width=0.5\textwidth]{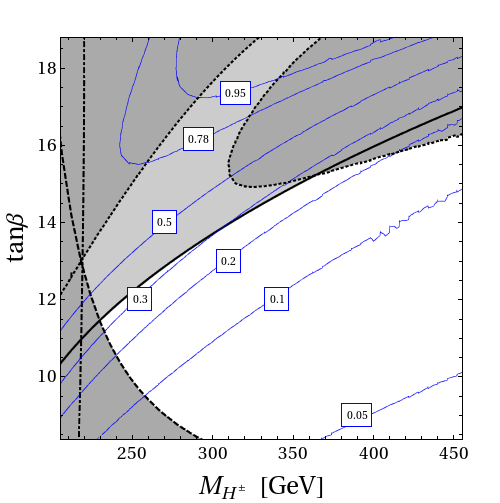}
\caption{
Possible values of $S_{\psi\phi}$ in the $M_{H^\pm}$ - $\tan\beta$ plane for the example scenario described in the text.
The dark gray region is excluded by the constraints from $\Delta M_s$ (dotted black line), from BR$(B \to X_s \gamma)$ (dashed-dotted black line) and the LEP bound on the lightest Higgs mass (dashed black line). The light gray region is excluded by the data on BR$(B_s \to \mu^+ \mu^-)$ (solid black line) only.
}
\label{fig:MHp_tanb_Spsiphi}
\end{figure}

In Fig.~\ref{fig:MHp_tanb_Spsiphi} we show an example scenario capable of producing a large $B_s$ mixing phase and being generically in agreement with the experimental constraints on $M_{H_1}$, $\Delta M_s$, BR$(B_s\to\mu^+\mu^-)$ and BR$(B \to X_s\gamma)$. We fix $|\omega| = 0.4$, $|\alpha \omega| = 2$, Arg$(\omega) = - 0.75$, Arg$(\alpha) = - 2$, $\mu = -950$~GeV, $m_S = 1000$~GeV, $M = 6$~TeV, 3rd generation squark soft masses $m_{\tilde t} = m_{\tilde b} = 500$~GeV, 1st and 2nd generation squark masses $\tilde m = 4$~TeV, slepton masses $\tilde m_\ell = 4$~TeV, trilinear couplings $A_t = - 2.5 m_{\tilde t}$, gaugino masses $M_1 = 200$~GeV, $M_2 = 400$~GeV, $M_3 = 1200$~GeV. We plot contours of constant $S_{\psi\phi}$ in the $M_{H^\pm}$ - $\tan\beta$ plane. 

As expected, the largest values for $S_{\psi\phi}$ that are in agreement with the BR$(B_s \to \mu^+ \mu^-)$ bound are obtained for a rather light Higgs spectrum ($M_{H^\pm}\sim 240$~GeV) and moderate values of $\tan\beta \simeq 11$. Values for the $B_s$ mixing phase up to $S_{\psi\phi} \simeq 0.3$ can be reached in this example scenario.

\begin{figure}[t] \centering
\includegraphics[width=0.45\textwidth]{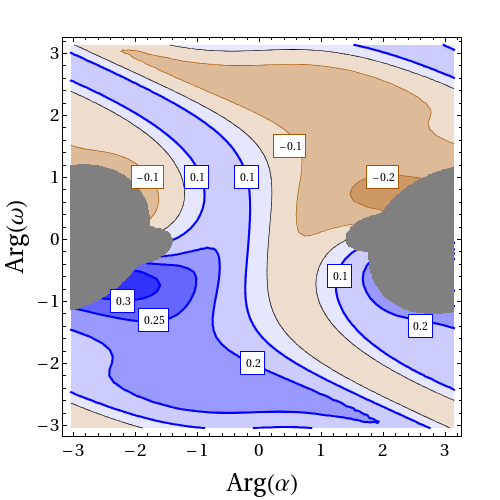} ~~~
\includegraphics[width=0.45\textwidth]{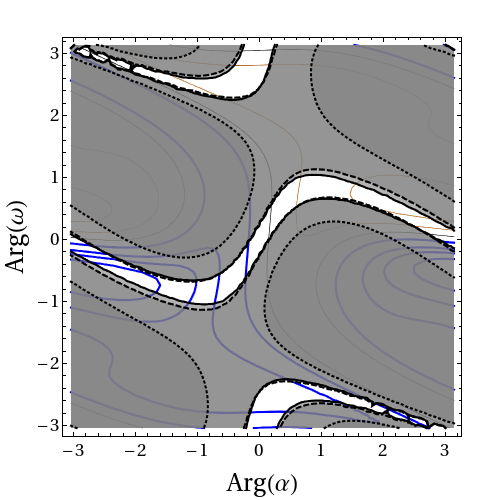}
\caption{
Left: Possible values of $S_{\psi\phi}$ in the Arg$(\alpha)$ - Arg$(\omega)$ plane for the example scenario of Fig.~\ref{fig:MHp_tanb_Spsiphi} (see text for details). The dark gray region is excluded by the constraint from $\Delta M_s$, the LEP bound on the lightest Higgs mass and the experimental data on BR$(B_s \to \mu^+ \mu^-)$ and BR$(B \to X_s \gamma)$.
Right: Same plot as on the left, but with constraints from EDMs. The solid, dashed and dotted black lines correspond to the Thallium, Mercury and neutron EDMs respectively.
}
\label{fig:alpha_omega_Spsiphi}
\end{figure}

The chosen large phases for $\omega$ and $\alpha$ together with the light Higgs spectrum and the not very small $\tan\beta$ will generically also induce huge contributions to EDMs. 
In Fig.~\ref{fig:alpha_omega_Spsiphi} we show the value of $S_{\psi\phi}$ in the Arg$(\alpha)$ - Arg$(\omega)$ plane, fixing $\tan\beta$ and $M_{H^\pm}$ to 11 and 240~GeV, respectively and all the other parameters chosen as in the scenario presented above.
While in the left plot only the constraints from $\Delta M_s$, the lightest Higgs mass, BR$(B \to X_s \gamma)$ and BR$(B_s \to \mu^+ \mu^-)$ are applied, the right plot shows the regions that are excluded by the EDMs. 
We observe that even though the EDMs severely restrict the allowed values for Arg$(\alpha)$ and Arg$(\omega)$, there are regions in parameter space with sizable $S_{\psi\phi}$ that pass the constraints from the EDMs. This happens because the contributions to the $B_s$ mixing amplitude and the contributions to the EDMs depend on different combinations of the phases Arg$(\alpha)$ and Arg$(\omega)$.

We remark that the Thallium, Mercury and neutron EDMs in principle also depend on different combinations of the NP phases because of their different sensitivity to the 1- and 2-loop contributions. For not very small $\tan\beta$ the different EDMs tend to generically give complementary constraints in the Arg$(\alpha)$ - Arg$(\omega)$ plane and largely exclude non-trivial phases. This does not happen in the example scenario above where we chose heavy 1st and 2nd generation squark as well as heavy sleptons. In that way the 1-loop contributions to the EDMs decouple to a large extent and the Thallium, Mercury and neutron EDM are all dominated by very similar 2-loop contributions, leading to approximately aligned constraints in the Arg$(\alpha)$ - Arg$(\omega)$ plane. 

We also remark that the right plot of Fig.~\ref{fig:alpha_omega_Spsiphi} reflects the situation where Arg$(\alpha)$ and Arg$(\omega)$ are the only CP violating phases of the model. If also the $\mu$ term or the soft SUSY breaking parameter introduce CP violation, additional 1-loop contributions to EDMs arise. Large cancellations among the several contributions can then in principle occur and the parameter space for Arg$(\alpha)$ and Arg$(\omega)$ opens up. An extensive analysis of the model, also allowing for a complex $\mu$ term and a complex soft SUSY breaking sector is however beyond the scope of this work.

\begin{figure}[t] \centering
\includegraphics[width=0.48\textwidth]{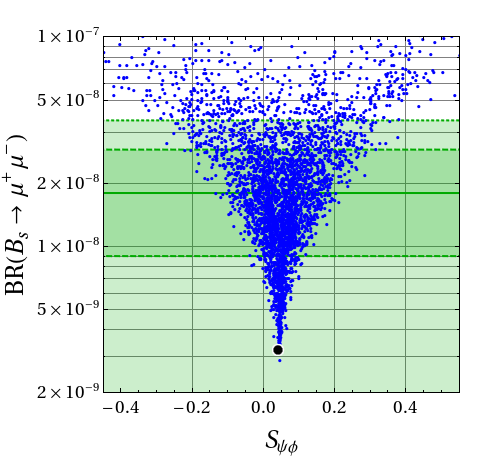} ~~~
\includegraphics[width=0.45\textwidth]{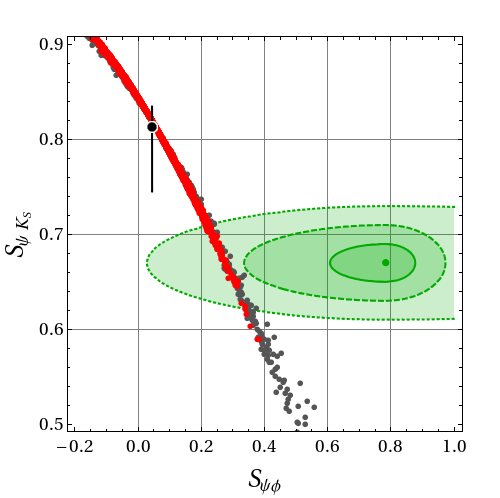}
\caption{
Left: Correlation between $S_{\psi\phi}$ and BR$(B_s \to \mu^+\mu^-)$. The solid green line represents the current central value of the CDF result~\cite{Collaboration:2011fi} on BR$(B_s \to \mu^+\mu^-)$, while the dashed lines correspond to the experimental $1\sigma$ range. The dotted line finally shows the 95\% C.L. bound.
Right: Correlation between $S_{\psi\phi}$ and $S_{\psi K_S}$. The green areas show the fit results for $S_{\psi\phi}$ from~\cite{Lenz:2010gu} combined with the experimental measurement of $S_{\psi K_S}$ at the 1, 2 and 3$\sigma$ level. Red (light gray) points are allowed by all constraints, (dark) gray points are excluded by BR$(B_s \to \mu^+\mu^-)$.
In both plots the black point shows the central SM values. In the right plot the vertical black lines indicates the $1\sigma$ uncertainty of the SM prediction of $S_{\psi K_S}$.
}
\label{fig:Spsiphi_scatter}
\end{figure}

Having established that there are indeed regions in parameter space where a sizable $S_{\psi\phi}$ is possible, we show in Fig.~\ref{fig:Spsiphi_scatter} the results of a parameter scan of the model. 
We keep the gaugino, slepton and 1st two generation squark masses as well as the scale $M$ to the values of the example scenario defined above and allow the remaining parameters to vary in the ranges
$|\omega| < 1$, $|\alpha \omega| < 2$, $- \pi <$ Arg$(\omega)$, Arg$(\alpha) < \pi$, $-1000$~GeV $< \mu < -500$~GeV, $500$~GeV $< m_S < 1000$~GeV, $500$~GeV $< m_{\tilde t} = m_{\tilde b} < 1000$~GeV, $A_t = - (2-3) m_{\tilde t}$, $\tan\beta < 20$ and $M_{H^\pm} < 500$~GeV.
The left plot of Fig.~\ref{fig:Spsiphi_scatter} shows the correlation between $S_{\psi\phi}$ and BR$(B_s \to \mu^+ \mu^-)$. We observe that in our framework the current experimental data on BR$(B_s \to \mu^+ \mu^-)$ gives an absolute limit on the $B_s$ mixing phase of $S_{\psi\phi} \lesssim 0.4$\footnote{However, if we assume the absence of the additional operators arising at the $1/M^2$ level that can stabilize the EW minimum, then the requirement of an absolute stable EW minimum implies $S_{\psi\phi} \lesssim 0.1$.}, with the central value for BR$(B_s \to \mu^+ \mu^-)$ leading to $S_{\psi\phi} \lesssim 0.25$.
Furthermore, for any given value of $S_{\psi\phi}$, the model predicts a lower bound on BR$(B_s \to \mu^+ \mu^-)$. For $S_{\psi\phi} \simeq 0.2$ for example we obtain BR$(B_s \to \mu^+ \mu^-) \gtrsim 1.5 \cdot 10^{-8}$. Such values will be probed in the near future by LHCb~\cite{Aaij:2011rj}.

The right plot shows the strong correlation between $S_{\psi\phi}$ and $S_{\psi K_S}$ in the studied model. Large positive effects in $S_{\psi\phi}$ always imply sizable negative effects in $S_{\psi K_S}$ and vice versa, as it is expected because of their origin from $\tilde C_2$. The NP effects in the  $B_s$ mixing phase soften the tension between the SM prediction and the value for $S_{\psi\phi}$ preferred by recent fits~\cite{Ligeti:2010ia,Lenz:2010gu} and simultaneously the effects in $S_{\psi K_S}$ can lead to a very good agreement with its measurement.

\bigskip
We end this section by listing further predictions in the region of parameter space with largest $B_s$ mixing phase.
As it is evident from Fig.~\ref{fig:MHp_tanb_Spsiphi}, the mass of the lightest Higgs is close to the LEP bound of $114.4$~GeV while the two heavier Higgs bosons have masses of $M_{H_{2,3}} \simeq (200 - 300)$~GeV.
Despite the rather large mass splitting, the lightest Higgs does contain non-negligible components of the non-SM like Higgs bosons in that region of parameter space and therefore has enhanced couplings to $\tau \bar\tau$ and in particular to $b \bar b$, due to the $\tan\beta$ resummation factors. Correspondingly, its branching ratio to $\gamma \gamma$ is strongly suppressed, such that observing this light Higgs boson at the LHC will be very challenging. On the other hand, due to the enhanced branching ratio into $b \bar b$ it might be possible to probe this Higgs in the $HV \to b \bar b$ channel, analyzing the full Tevatron data, as shown in Fig.~\ref{fig:collider_sensitivity}.
Also the two heavy Higgs bosons have enhanced couplings to bottoms and taus and they can be probed at the LHC in the inclusive $ H_{2,3} \to \tau \tau$ channels. The stops in this scenario are rather light with masses $m_{\tilde t} \lesssim 500$~GeV while the squarks of the first 2 generations and also the sleptons are expected to be very heavy in the multi TeV range.

All EDMs are generically predicted to be close to the current experimental bounds. However, no absolute lower bound can be put on the EDMs due to the possibility of large cancellations among different contributions.

Finally we mention that in SUSY frameworks where the CKM matrix is (effectively) the only source of flavor violation but additional CP phases are considered, visible effects in observables that are sensitive to CP violation in the $b \to s \gamma$ transition are also generically predicted~\cite{Altmannshofer:2008hc,Barbieri:2011vn}. Examples of such observables include the CP asymmetry in the $B \to X_s \gamma$ decay~\cite{Soares:1991te,Kagan:1998bh,Benzke:2010tq}, time dependent CP asymmetries in the $B \to \phi K_S$ and $B \to \eta^\prime K_S$ decays~\cite{Grossman:1996ke} as well as CP asymmetries in the $B \to K^* \ell^+ \ell^-$ decay~\cite{Bobeth:2008ij,Altmannshofer:2008dz}. However, in general no clear correlation with the $B_s$ mixing phase can be established, as these observables have different dependence on the SUSY parameters.

\section{Conclusions} \label{sec:conclusions}

In addition to the MSSM particle content, new degrees of freedom beyond the MSSM (BMSSM) might be present slightly above the TeV scale. In an effective field theory approach, the leading corrections from the BMSSM physics to the MSSM Higgs sector can be described by two dimension 5 operators. These two operators are a potential source of CP violation at the tree level. In this work we studied the impact of these new sources of CP violation beyond the MSSM on Higgs searches and B meson observables taking into account constraints from Electric Dipole Moments.
In the first part of the work we concentrated on the Higgs collider phenomenology that is specific of the BMSSM with CP violation. In the second part we analyzed a complementary region of parameter space where interesting effects in flavor physics are possible.

In contrast to the MSSM, the Higgs sector of the BMSSM generically is CP violating already at the tree level. Correspondingly the physical spectrum consists of one charged Higgs and three neutral Higgs bosons that are mixtures of the neutral scalar and pseudoscalar states.
Taking into account constraints from vacuum stability and Electric Dipole Moments as well as collider constraints from LEP and Tevatron, we worked out the Higgs collider phenomenology of the model. We discussed the Higgs spectrum and couplings as well as the dominant production cross sections and the Higgs branching fractions in several scenarios that show distinct features of the BMSSM with CP violation.

As long as $\tan\beta$ (as well as the BMSSM scale $M$) is not too large ($\tan\beta \lesssim 5$), the higher dimensional operators can significantly enhance the mass of the lightest neutral Higgs boson above the bound of $M_h \lesssim 135$~GeV that holds in the MSSM. As also EDM constraints become extremely strong for large values of $\tan\beta$, we concentrated on the low $\tan\beta$ regime in our analysis of the Higgs collider phenomenology.
The CP violating phases of the higher dimensional operators allow for sizable couplings of all three Higgs bosons to the weak gauge bosons. Correspondingly, we found that the most striking scenario that can be realized in the BMSSM with CP violation are three neutral Higgs bosons that all decay to the $WW$ final state. Such a scenario cannot be realized, either in the MSSM with CP violation or in the BMSSM without CP violation and is unique to the framework considered in this work. It can be probed at the 7~TeV run of LHC.

We also find that the model allows for benchmark scenarios where all three Higgs bosons are heavy with masses $M_{H_i} \gtrsim 150$~GeV but still all decay dominantly into $b \bar b$. Correspondingly, it will be challenging to probe such scenarios at the 7~TeV run of LHC as they predict signals in the studied Higgs search channels that are at the very border or even below the expected sensitivities with 10~fb$^{-1}$.

\bigskip\noindent
We also discussed distinct signals of the modified Higgs sector on flavor observables in a complementary region of parameter space, where the direct Higgs searches do not significantly depart from the MSSM expectations. In the large $\tan\beta$ regime, significant deviations from the MSSM predictions can in principle be expected in the $B_{d,s} \to \mu^+ \mu^-$ decays and in $B_{d,s}$ mixing. Such processes can be affected by NP contributions from Higgs and double Higgs penguins and are therefore highly sensitive to the Higgs spectrum and couplings.
We find that the main qualitative difference with respect to the MSSM are NP contributions to the $\Delta F = 2$ operator $(\bar b P_L q)^2$~. In the MSSM this operator is highly suppressed by $\cot^2\beta ~M_W^2/M_A^2$ and completely negligible. Instead, the leading double Higgs contributions to the meson mixing amplitudes in the MSSM generate the operator $(\bar b P_R q)(\bar b P_L q)$. Assuming Minimal Flavor Violation (MFV), i.e. the absence of new sources of flavor violation in addition to the CKM matrix, these contributions are proportional to $m_b m_s$ and $m_b m_d$ in the case of $B_s$ and $B_d$ mixing respectively. Such contributions are strongly constrained by the upper bound on the branching ratio of the $B_s \to \mu^+ \mu^-$ decay in the MSSM with MFV and therefore CP violation in meson mixing remains SM like. Correspondingly the hints from Tevatron towards a large $B_s$ mixing phase cannot be addressed in the MSSM with MFV. 

In the BMSSM with MFV on the other hand, the $(\bar b P_L q)^2$ operator typically gives the dominant contribution. It is proportional to $m_b^2$ both for $B_s$ and $B_d$ mixing and can induce large CP phases to both mixing amplitudes that are equal in size and phase. 
Even though the $B_s \to \mu^+ \mu^-$ decay still gives the strongest constraint on double Higgs penguin contributions to meson mixing, we find that a $B_s$ mixing phase up to $S_{\psi\phi} \lesssim 0.4$ can be achieved in the BMSSM with MFV, compatible with EDM constraints.
We find that stability of the electro-weak vacuum in the region of parameter space with a large $B_s$ mixing phase requires the presence of additional $1/M^2$ operators with appropriately chosen coefficients.
In addition, a large $B_s$ mixing phase of $S_{\psi\phi} \simeq 0.4$ implies a sizable suppression of the SM prediction for the $B_d$ mixing phase $S_{\psi K_S}$, which is welcome in view of the observed tensions in the determination of the unitarity triangle. Simultaneously, also the BR$(B_s \to \mu^+ \mu^-)$ is predicted close to the upper bound of the region recently reported by CDF. Interestingly, a sizable $S_{\psi\phi} \gtrsim 0.25$ implies a lower bound on BR$(B_s\to\mu^+\mu^-) \gtrsim 2 \times 10^{-8}$ which is just above the central value given by CDF.

To conclude, in this paper we have shown that the BMSSM with CP violation can lead to complementary benchmark scenarios that either show novel signatures in Higgs collider physics or can ameliorate some tensions in present B physics data. The predicted signals are characteristic of the presence of NP phases in the Higgs potential and allow to distinguish the model both from the CP violating MSSM and the BMSSM without CP violation and will be tested in the near future at the LHC.

\section*{Note Added}

After the completion of this work, new Tevatron and LHC bounds on Higgs searches~\cite{TevatronEPS,AtlasEPS,CMSEPS} and on the branching ratio of the decay $B_s \to \mu^+ \mu^-$~\cite{LHCb-CONF-2011-037,Chatrchyan:2011kr,CMS-PAS-BPH-11-019} appeared.

The ATLAS collaboration presented the combined exclusion limit for a SM Higgs with an integrated luminosity ranging from 1.04 and 1.21 $\rm{fb}^{-1}$~\cite{AtlasEPS}. The CMS collaboration presented the combination of six SM Higgs boson searches corresponding to 1.0-1.1 $\rm{fb}^{-1}$ of integrated luminosity~\cite{CMSEPS}. Our Higgs scenarios lie beyond the range presently probed by the LHC, even if the lightest Higgs boson of the first two scenarios will be soon tested by the LHC data. We observe that, for Higgs masses below $\sim 180$ GeV, the observed limits of ATLAS and CMS are weaker than the projected limits we presented in Fig.~\ref{fig:collider_sensitivity}.

Concerning the branching ratio of the rare decay $B_s\to \mu^+ \mu^-$, a preliminar combination has been performed using the LHCb bound obtained with 337 $\rm{pb}^{-1}$ of integrated luminosity ($1.5\times10^{-8}$) and the CMS bound obtained with 1.14 $\rm{fb}^{-1}$ of integrated luminosity ($1.9\times10^{-8}$). The combined value is BR$(B_s \to \mu^+ \mu^-)<1.1\times 10^{-8}$ at the 95\% C.L.~\cite{CMS-PAS-BPH-11-019}. This result does not confirm the excess observed by CDF~\cite{Collaboration:2011fi}. Using this updated bound on the decay of the $B_s$ meson, we find that the possible range for the $B_s-\bar B_s$ mixing phase is $S_{\psi\phi}\leq0.15$ (as observed from Fig.~\ref{fig:Spsiphi_scatter}). This value is however considerably larger than the one possible in the MSSM with MFV.

\section*{Acknowledgments}

We thank M.~Neubert, E.~Ponton and C.E.M~Wagner for insightful comments and A.~Delgado, Y.~Grossman and J.~Zurita for interesting discussions.
Fermilab is operated by Fermi Research Alliance, LLC under Contract No. De-AC02-07CH11359 with the United States Department of Energy.

\begin{appendix}
\section{Chargino, Neutralino and Squark Masses}\label{app:SUSY_masses}

In this appendix we summarize the impact of the $1/M$ suppressed operators in the Higgs sector on the chargino, neutralino and squark masses.

We decide to factor out the phases $\theta_u$ and $\theta_d$ of the fields $H_u$ and $H_d$  (see~(\ref{eq:Higgs_fields})) also on the Higgsino fields $\tilde H_{u,d}$, i.e. we factor out the phases on the entire Higgs superfields $\hat H_{u,d}$.
In this way the terms that mix the Higgsinos and gauginos after electro-weak symmetry breaking are real and the physical Higgs phase $\theta$ appears only in the Higgsino part of the chargino and neutralino mass matrices.
In particular, the $2\times 2$ chargino mass matrix reads
\begin{equation}
M_{\chi^\pm} = \begin{pmatrix} M_2 & \frac{g_2}{\sqrt{2}} v s_\beta \\ \frac{g_2}{\sqrt{2}} v c_\beta & \mu e^{i\theta} - \omega_1 \frac{v^2}{M}s_\beta c_\beta e^{2i\theta}  \end{pmatrix}~.
\end{equation}
and the $4\times 4$ neutralino mass matrix is given by
\begin{equation}
M_{\chi^0} = \begin{pmatrix} 
M_1 & 0 & -\frac{g_1}{2} v c_\beta & \frac{g_1}{2} v s_\beta \\ 
0 & M_2 & \frac{g_2}{2} v c_\beta & -\frac{g_2}{2} v s_\beta \\
-\frac{g_2}{2} v c_\beta & \frac{g_2}{2} v c_\beta & \omega_1 \frac{v^2}{M} c_\beta^2 e^{2i\theta} & -\mu e^{i \theta} + 2\omega_1 \frac{v^2}{M} s_\beta c_\beta e^{2i\theta}\\
\frac{g_2}{2} v s_\beta & -\frac{g_2}{2} v s_\beta & -\mu e^{i \theta} + 2\omega_1 \frac{v^2}{M} s_\beta c_\beta e^{2i\theta} & \omega_1 \frac{v^2}{M} s_\beta^2 e^{2i\theta}  \\
\end{pmatrix}~.
\end{equation}
Given the non-zero phase of the Higgs VEVs we also perform a phase shift on the right-handed quark and lepton fields in order to have real Yukawas at the tree level. As in case of the Higgs fields, we apply these phase shifts on the entire quark and lepton superfields. In that way the left-right mixing terms in the squark and slepton mass matrices only contain the physical phase $\theta$ and not $\theta_u$ and $\theta_d$ separately.
The up and down squark masses read
\begin{equation}
M_u^2 = \begin{pmatrix} \tilde m^2 + m_u^2 - \frac{c_{2\beta}}{6}(M_Z^2-4M_W^2) & -m_u (A_u + \mu^* /t_\beta e^{-i\theta} - \omega_1^* \frac{v^2}{M} c_\beta^2 e^{-2i\theta}) \\ -m_u (A_u^* + \mu /t_\beta e^{i\theta} - \omega_1 \frac{v^2}{M} c_\beta^2 e^{2i\theta}) & \tilde m^2 + m_u^2 + \frac{2c_{2\beta}}{3} M_Z^2 s_W^2\end{pmatrix} ~,
\end{equation}
\begin{equation}
M_d^2 = \begin{pmatrix} \tilde m^2 + m_d^2 - \frac{c_{2\beta}}{6}(M_Z^2+2M_W^2) & - m_d (A_d + \mu^* t_\beta e^{-i\theta} - \omega_1^* \frac{v^2}{M} s_\beta^2 e^{-2i\theta}) \\ - m_d (A_d^* + \mu t_\beta e^{i\theta} - \omega_1 \frac{v^2}{M} s_\beta^2 e^{2i\theta}) &  \tilde m^2 + m_d^2 -\frac{c_{2\beta}}{3} M_Z^2 s_W^2 \end{pmatrix}~,
\end{equation}
where, for simplicity, we assumed a common mass $\tilde m$ for the left-handed squark doublets and the right-handed squark singlets.

\section{Electroweak Precision Constraints}\label{app:ST}
 
In this appendix we report the expressions for the contributions to the S and T parameter~\cite{Peskin:1991sw} from the Higgs sector of the BMSSM with CP violation. Generalizing the results given in~\cite{He:2001tp,Carena:2009gx} to our case with CP violation we find
\begin{eqnarray}
\pi M_Z^2 S &=& \sum_{ijk} \xi_{ZZH_i}^2 \frac{1}{2} \epsilon_{ijk}^2 \Big[ \mathcal{B}_{22}(M_Z^2,M_{H_j}^2,M_{H_k}^2) -\mathcal{B}_{22}(M_Z^2,M_{H^\pm}^2,M_{H^\pm}^2)  \Big] \nonumber \\
&& + \sum_i \xi_{ZZH_i}^2 \Big[ \mathcal{B}_{22}(M_Z^2,M_Z^2,M_{H_i}^2) - \mathcal{B}_{22}(M_Z^2,M_Z^2,M_\textnormal{ref}^2) \nonumber \\
&& ~~~~~~~~~~~ - M_Z^2 \mathcal{B}_0(M_Z^2,M_Z^2,M_{H_i}^2) + M_Z^2 \mathcal{B}_0(M_Z^2,M_Z^2,M_\textnormal{ref}^2) \Big] ~,
\end{eqnarray}
\begin{eqnarray}
16 \pi s_W^2 M_W^2 T &\!\!\!=\!\!\!& 3 \sum_i \xi_{ZZH_i}^2 \Big( G(M_{H_i}^2,M_Z^2) - G(M_\textnormal{ref}^2,M_Z^2) - G(M_{H_i}^2,M_W^2) + G(M_\textnormal{ref}^2,M_W^2) \Big) \nonumber \\
&& + \sum_i |\xi_{WHH_i}|^2 G(M_{H_i}^2,M_{H^\pm}^2) - \sum_{ijk} \xi_{ZZH_i}^2 \frac{1}{2} \epsilon_{ijk}^2 G(M_{H_j}^2,M_{H_k}^2)~.
\end{eqnarray}
For the loop function $G$ one has 
\begin{equation}
G(x,y) = \frac{1}{2}(x+y) - \frac{xy}{x-y}\log\left(\frac{x}{y}\right) 
\end{equation}
and $\mathcal{B}_{22}$ and $\mathcal{B}_0$ can be found in~\cite{He:2001tp}. The mass $M_\textnormal{ref}$ represents a reference value for the Higgs boson mass in the SM.
The effective couplings of the Higgs bosons with two gauge bosons, $\xi_{ZZH_i}$, and the effective couplings of the Higgs bosons with the charged Higgs boson and a $W$, $\xi_{WHH_i}$, that enter the above expressions, are given by
\begin{eqnarray}
\xi_{WHH_i} &=& s_{\beta - \alpha} O_{2i} - c_{\beta-\alpha} O_{1i} - i O_{3i} ~, \\
\xi_{ZZH_i} &=& s_{\beta - \alpha} O_{1i} + c_{\beta-\alpha} O_{2i} .
\end{eqnarray}
We explicitly checked that all the scenarios considered in this work are compatible with the constraints from S and T.

\section{Loop Functions}\label{app:loop}

\begin{eqnarray}
f_d(x) &=& \frac{4(1+5x)}{9(1-x)^3} + \frac{8x(2+x)\log x}{9(1-x)^4} ~, \\
\tilde f_d(x) &=& -\frac{2(x+11)}{3(1-x)^3} + \frac{(x^2-16x-9)\log x}{3(1-x)^4} ~, \\
f_e(x,y) &=& \frac{7y-x(y-7)-13}{4(1-x)^2(1-y)^2} + \frac{(2+x)\log x}{2(1-x)^3(y-x)} + \frac{(2+y)\log y}{2(1-y)^3(x-y)} ~, \\[12pt]
f(z) &=& \int_0^1 dx \frac{1-2x(1-x)}{x(1-x)-z}	\log\left( \frac{x(1-x)}{z} \right)~, \\[12pt]
f_1(x) &=& \frac{1}{1-x} + \frac{x}{(1-x)^2} \log x ~, \\
f_2(x,y)&=& \frac{x \log x}{(1-x)(y-x)} + \frac{y \log y}{(1-y)(x-y)}~, \\[12pt]
h_7(x) &=& - \frac{5x^2-3x}{12(1-x)^2} - \frac{3x^2-2x}{6(1-x)^3}\log{x}~, \\
h_8(x) &=& -\frac{x^2-3x}{4(1-x)^2} + \frac{x}{2(1-x)^3}\log{x}~,\\
f_7(x) &=& - \frac{13-7x}{24(1-x)^3} - \frac{3+2x-2x^2}{12(1-x)^4}\log{x}~, \\
f_8(x) &=& \frac{1+5x}{8(1-x)^3} + \frac{x(2+x)}{4(1-x)^4}\log{x}~.
\end{eqnarray}

\end{appendix}

\bibliographystyle{utphys}
\bibliography{mybib}


\end{document}